\documentclass[aps,pre,showkeys,amssymb]{revtex4-2}
\usepackage{graphicx}
\usepackage{color}
\usepackage[utf8]{inputenc}
\usepackage{braket}
\usepackage{graphicx}
\usepackage{appendix}
\usepackage{adjustbox}

\begin{document}

\title{Predicting temperatures in Brazilian states capitals via Machine Learning}

\author{Sidney T. da Silva$^1$, 
Enrique C. Gabrick$^{2,}$\footnote{Corresponding Author \\e-mail: ecgabrick@gmail.com}, 
Ana Luiza R. de Moraes$^2$, 
Ricardo L. Viana$^3$, 
Antonio M. Batista$^{4,5}$, 
Iber\^e L. Caldas$^2$, 
J\"urgen Kurths$^{6,7}$}
\affiliation{
$^{1}$Federal University of Paran\'a, 81531-980, Curitiba, PR, Brazil.\\
$^{2}$Institute of Physics, University of S\~ao Paulo, 05508-090, S\~ao Paulo, SP, Brazil. \\
$^3$Federal University of Paran\'a, Department of Physics and Interdisciplinary Center for Science, Technology and Innovation, Center for Modeling and Scientific Computing, 81531-980, Curitiba, PR, Brazil. \\
$^4$Graduate Program in Science, State University of Ponta Grossa, 84030-900, Ponta Grossa, PR, Brazil.\\
$^5$Department of Mathematics and Statistics, State University of Ponta Grossa, 84030-900, Ponta Grossa, PR, Brazil.\\
$^6$Potsdam Institute for Climate Impact Research, Telegrafenberg A31, 14473 Potsdam, Germany.\\
$^7$Department of Physics, Humboldt University Berlin, Newtonstra{\ss}e 15, 12489 Berlin, Germany.  
}

\begin{abstract}
Climate change refers to substantial long-term variations in weather patterns.
In this work, we employ a Machine Learning (ML) technique, the 
Random Forest (RF) algorithm, to forecast   
the monthly average temperature for  Brazilian's states 
capitals (27 cities) and the whole country, from January 1961 until December 2022.  
To forecast the temperature at $k$-month, we consider as features in RF:
$i)$ global emissions of carbon dioxide (CO$_2$), methane (CH$_4$), 
and nitrous oxide (N$_2$O) at $k$-month; 
$ii)$ temperatures from the previous three months, i.e., $(k-1)$, $(k-2)$ and 
$(k-3)$-month; 
$iii)$ combination of $i$ and $ii$.  
By investigating breakpoints in the times series,  
we discover that 24 cities and the gases present breakpoints in the 80's and 90's. 
After the breakpoints, we 
find an increase in the temperature and the gas emission.
Thereafter, we separate the cities according to their geographical position 
and employ the RF algorithm to forecast the temperature from 2010-08 until 2022-12. 
{
Based on $i$, $ii$, and $iii$, we find that the three inputs result in 
a very precise forecast, with a normalized root mean squared error (NMRSE)  
less than 0.083 for the considered cases. 
From our simulations, the better forecasted region is 
Northeast through $iii$ (NMRSE = 0.012).} 
Furthermore, we also investigate the forecasting of anomalous temperature data by removing 
the annual component of each time series.
{ In this case, 
the best forecasting is obtained with strategy $i$, with the best region being 
Northeast (NRMSE = 0.090).}
\end{abstract}

\maketitle

\section{Introduction}
Climate change has emerged as one of our time's most important problems 
and debated issues \cite{IPCC2023}. The general scientific consensus is 
that the Earth is warming \cite{oreskes04}. The temperature increases 
are responsible for changing weather patterns, which also affects  
biodiversity \cite{Cheng2013}. The implications of a global warming extend far beyond
environmental concerns \cite{hulme99}, affecting the global economy, public
health \cite{Souza2024}, food security \cite{wheeler13}, and ultimately, the future of our
civilization. 

Many studies have reported evidences about a planet warming at an
alarming rate \cite{Hansen2006}. 
One of the reasons for the warming is associated with  
human activities, including greenhouse gases and aerosol emissions \cite{Gillet2021}. 
Meanwhile, the contribution of natural forcings is smaller compared 
to the previous reasons \cite{Gillet2021}.   
The anthropogenically caused global climate change is responsible 
for decreasing the occurrence of cold days and increasing the frequency of 
warm days \cite{Alexander2006}. 
One of the significant contributors to this warming are the greenhouse
gases (GHGs), such as carbon dioxide (CO$_2$), methane (CH$_4$), and nitrous oxide
(N$_2$O) \cite{lashof90}. These gases are produced by fossil fuel combustion,
industrial activities, deforestation, among others. 
Due to the correlation between the increase in gas emissions and the temperature, 
GHGs can be used to predict temperature increasing \cite{crowley00,Vu2024}.

In 
recent years, many researches have analysed the impacts and 
forecast global climate changes \cite{whitmarsh11,Bennet2023,Wunderling2021,Karl2003,Tang2013,Rosenzweig2008}. 
As a particular case, many scholars have investigated the South American 
climate pattern \cite{Grimm2000,Haylock2005,Marengo2011}. 
The impacts of global warming are diverse, for example, several  
scenarios suggest potential socioeconomic effects in Brazil due to the 
impacts of climate change on agriculture. An analysis of these scenarios 
showed that Center-West and Northeast will be more affected than other 
Brazilian regions \cite{Santos2022}. 
Confalonieri et al. \cite{Confalonieri2009} 
conducted a study to measure the public health vulnerability in Brazil due 
to effects of climate change. They found that the Northeast is 
the most vulnerable Brazilian region. As observed by Chagas et al. \cite{Chagas2022}, 
the use of water and deforestation  have contributed to the increase 
of climate changes effects. In 42\% of the studied area, they showed that the 
drying is related to the decrease of rainfall and increase of water use 
in agricultural zones. On the other hand, in 29\% of the studied area, they 
found that the increase of severe floods and droughts is related to more 
extreme rainfall and deforestation.  Then, climate change is crucial for 
the future of agriculture (familiar or in large scale) and food security \cite{Rattis2021}.

From a modelling perspective, some works have studied data and predict new scenarios. 
Marengo et al. \cite{Marengo2009} 
considered the PRECIS regional climate modelling to analyse and forecast 
the temperature and precipitation in South America, in historical and projected 
data. They reported that the occurrence of warm nights 
will be more frequent in the future. Furthermore, they observed significant changes in rainfall 
and dry spells in South America, showing a 
intense change in the frequency of dry days, which becomes even more frequent. 
The increasing trend of warm nights was also observed 
by Vincent et al. \cite{Vincent2005}, by analysing data from South America. 
Ballarin et al. \cite{Ballarin2023} organised a data set based 
on General Circulation and Earth System Models  to study the historical 
(1980-2013) and projected future (2015-2100) climate changes in Brazil. 
In addition to these modelling technique, which are mostly based on statistical 
models and physics based, also the to use of complex networks frameworks \cite{Feldhoff2015,Bosikun2025}  
or Machine Learning (ML) based models \cite{zhong21} are promising.

Among the different forms of modelling, ML techniques have been 
shown significant accuracy in predictions \cite{Bracco2025}. 
Zennaro et al. \cite{zennaro21} presented a systematic review about ML 
for risk assessment due to climate change. They analysed the publications from 2000 until 2020, where many ML techniques 
were employed to explore and forecast risk assessment. 
 In another work, 
V\'azquez-Ram\'irez et al. \cite{ramirez23} proposed a finite-time thermodynamic 
model to study and predict the increase of surface temperature. 
They compared their method with results based on Linear 
Regression, Ridge Regression, and Artificial Neural Networks. 
Considering emissions of CO$_2$,  N$_2$O, 
and  CH$_4$. Zheng \cite{zheng18} employed several ML methods to 
verify the warming of the Earth. Among these methods, the author used  
Linear Regression, Lasso, Support Vector Regression, and Random Forest (RF). 
In a comparative study, Zheng demonstrated that RF is more accurate to forecast 
the temperature than the other considered models. The author developed a 
precise forecasting of temperature by using the gas emission 
as feature in the RF algorithm. In addition, ML algorithms can be employed to study the impacts 
on marine ecosystems, as studied by  
Alhakami et al. \cite{alhakami22}. 
They reported that fish stocks
are affected in a scenario of rapid global warming. 
In addition to the previous discussed works, other researchers have been 
exploring ML techniques to forecast effects of climate changes, e.g., 
Refs. \cite{Sanz2016,Orsenigo2018,Ustaoglu2008}.

This work's main goal is to study temperature changes in Brazil and their predictions.
To do that, we use a data set from 
capitals of Brazilian states and  propose three strategies as features in the 
RF algorithm, a ML technique, to forecast the monthly temperatures in 
these capitals.  
Firstly, we consider the  
monthly average temperature from 27 Brazilian's states capitals from 1961-01 until 
2022-12 and  explore the statistical properties of 
their time series. We mainly find that 2 capitals exhibit breakpoints 
in the 70's, 18 in the 80's, 6 in the 90's, and just 1 has a breakpoint in the 21st century. 
The breakpoints are associated with a change in the average temperature 
and average trend from the time series. We verify that these 
properties increase for all the analyzed cases. By grouping these capitals 
according to their geographical region, we uncover that South, Southeast, and 
Northeast time series have breakpoints in the 80's, while Center-West and 
North in the 90's. The breaks occurred for the gases in the  
80's for CH$_4$ and in the 90's for CO$_2$ and N$_2$O. 
In addition to these 
time series, we consider the temperature for the whole of Brazil, which 
changed its properties in 2001. We apply three input strategies in the RF algorithm. As target,  
we consider the monthly average temperature and anomalous temperature data, 
at $k$-month. 
The strategies are: $i)$ global emissions of CO$_2$, CH$_4$, 
and N$_2$O, at $k$-month; 
$ii)$ temperatures from the previous three months, i.e., $(k-1)$, $(k-2)$, 
$(k-3)$-month; 
$iii)$ combination of $i$ and $ii$. We demonstrate that the three strategies 
are appropriated to forecast the temperature for the normal data, while 
for the anomalous data, the best one is $i$. 
{In the case of normal data, it is not possible to choose 
the best candidate, once one set predicts better than others a given 
region but fails in another. Therefore, a significant contribution 
of this paper is to show the non-existence of a global set of features, 
but to show that the system is very complex and the inputs need to be 
particularly examined in order to improve the forecast. }

The manuscript is structured as follows: in Section \ref{metodos}, we describe the RF method
as the acquisition and processing of data. Section \ref{time_series_analysis} is devoted
to time series analysis, and Section \ref{ml_section} exhibits the predictions 
using the ML technique. Finally, our conclusions are drawn in
Section \ref{conclusao}.
\section{Methods} \label{metodos}
\subsection{Data acquisition and processing}
\begin{figure}[htbp]
\centering
\includegraphics[scale=0.4]{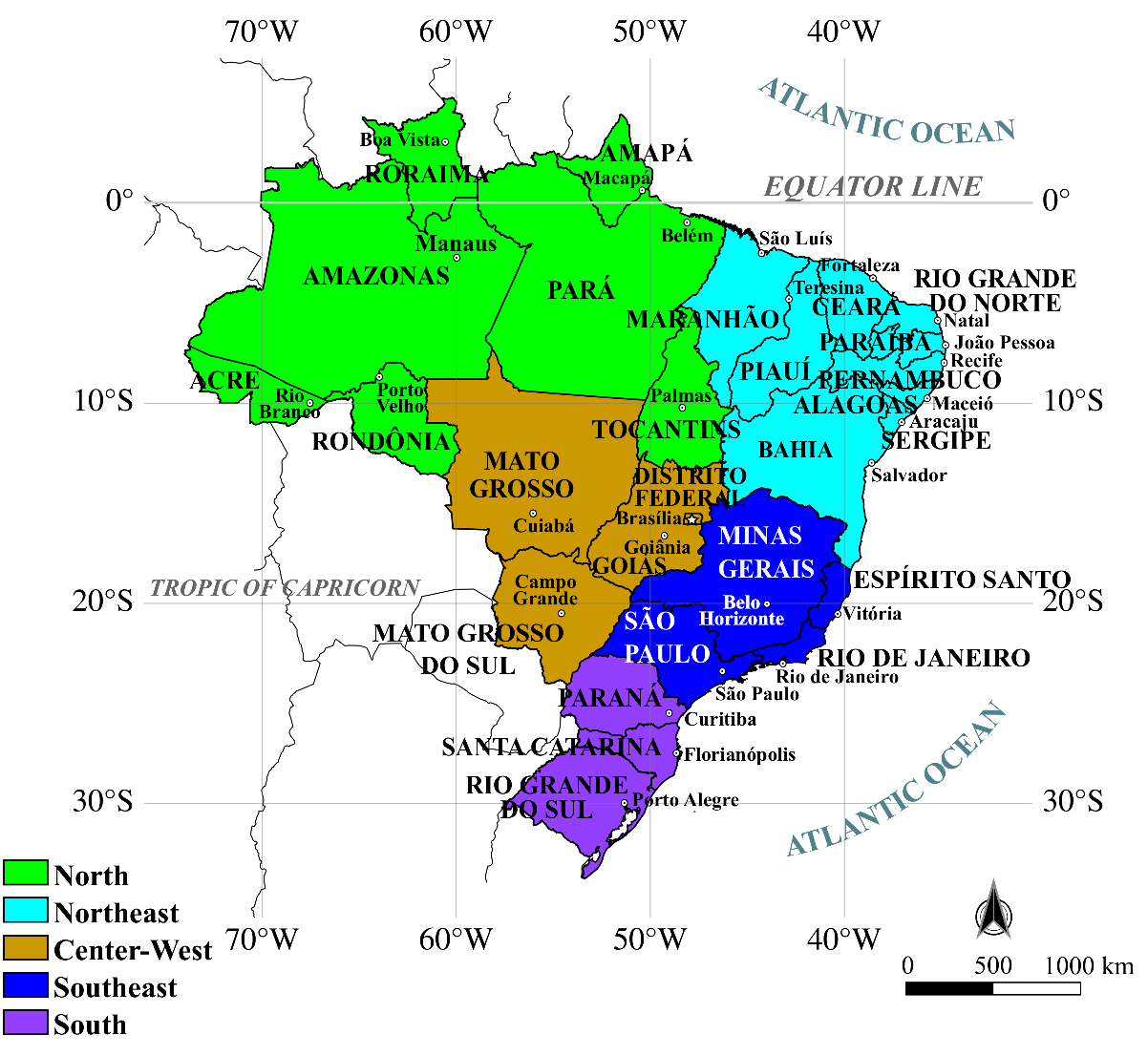}
\caption{Map of Brazil with respective states and their capitals. It is separated 
into five geographical regions: South (purple region), Southeast 
(blue region), Center-West (brown region), Northeast (cyan region), and 
North (green region).}
\label{fig1}
\end{figure}

In terms of geographical extension, Brazil is the largest country in South America and is geographically organised 
into 5 regions, namely South, Southeast, Center-West, Northeast, and North (Fig. \ref{fig1}). 
Each region is located in Fig. \ref{fig1} by purple, blue, 
brown, cyan and green colors, respectively, where the name of each state is in capital letters 
followed by the respective capitals, marked by a white circle in the map. 

As Brazil has 5700 cities, we analyse monthly average temperature of 
the Brazilian states capitals, totalling 27 time series from January 1961 
to December 2022. The dataset are obtained from  ERA5 reanalyses \cite{ERA5} 
and each time series is constituted by 744 elements. In order to reduce the number of 
time series, we average overall each region. For example, the time series 
that we call by South is the average overall Porto Alegre, Florian\'opolis, 
and Curitiba, and so on. The only exception is the Brazil time series, 
which is obtained by selecting the whole country in ERA5. In this way, 
we present the analyses of six time series and a detailed discussion about the 27 time series is in 
Appendix A. 

In addition to the dataset described above, we also consider the emission of greenhouse 
gases. We use the global emission of CO$_2$, CH$_4$, 
and N$_2$O, from January 1961 up to December 2022. The data is from the United States 
Environmental Protection Agency (EPA) \cite{gases}.

It is important to mention that all the dataset and codes employed in 
this research is available on  GitHub \cite{sidneygithub}.

\subsection{Breakpoints analysis}\label{breakpoints}
Considering each time series, we search for significant changes in their averages, 
variance, and standard deviations, called breakpoints analysis \cite{Cook2017}.  
We perform this analysis in Python, through the ruptures package \cite{ruptures} combined 
with the Dynamic programming method \cite{Tsitsiklis1996}. As  
a cost function, we consider the least absolute deviation \cite{Bai1995}. 

Before implementing the algorithm, we decompose the real time series ($y_t$) in 
terms of its components, i.e., seasonality ($s_t$), trend ($\tau_t$), and residuals ($\epsilon$). 
In this way, each time series is described by the follow mathematical form  
\begin{equation}
y_t = s_t + \tau_t + \epsilon_t, \label{components_eq}
\end{equation}
where the units are $^{\rm o}$C (degree Celsius). In general terms, the mathematical form 
of $s_t$ is a linear combinations of oscillatory functions, such 
as $s_t = b \, {\rm sin} (\omega_1 t + \phi_1) + c \, {\rm cos}(\omega_2 t + \phi_2)$, 
where $\omega_{1,2}$ are frequencies, $\phi_{1,2}$ phases, $b$ and $c$ 
constants. Meanwhile, the trend 
increases practically linearly with a deviation component ($\Gamma(t)$), 
given by $\tau_t = a + b \,t + \Gamma(t)$. The complete description of 
each component is in Section \ref{time_series_analysis}. 

After that, we implement the algorithm 
to identify the breakpoints. Once found, we calculate the averages ($\langle y \rangle$), 
variance  ($\sigma^2$) and standard deviation ($\sigma$) before and after 
the breaks. These quantities are compared to verify if there is significant 
change once crossed the breakpoints. In our analyses, we note that for 
all time series there is no change in the statistical properties 
($\langle y \rangle$, $\sigma^2$, and $\sigma$) of $s_t$. Therefore, 
in this work we discuss only changes in $y_t$ and $\tau_t$.

\subsection{Random Forest}
After the analysis previously discussed, we implement a ML technique 
to forecast the monthly average temperature. We adopt the 
Random Forest (RF) approach \cite{forest}, which is a supervised 
learning algorithm based on the Ensemble learning method 
and is composed of many decision trees \cite{Biau2016}.  
{ The justification for this choice is mostly based on the 
following criteria: as many trees are considered in the algorithm, the 
risk of overfitting is decreased and the accuracy is higher compared 
with individual models \cite{zheng18}. The applied algorithm does not require 
a linear relationship among the variables and can capture the nonlinearity 
from a given phenomenon \cite{Orsenigo2018}. For each considered feature, 
the algorithm returns its importance in the prediction \cite{GabrickDengue}.} 
In each decision tree, the choice 
of the partitions is made by considering 
the minimum mean squared error (MSE), given by 
\begin{equation}
{\rm MSE}_{min}=\min\left\{\sum_{i\in S_{k}}(y_i-{\langle y \rangle})^{2}\right\}, \label{eq1}
\end{equation}
where ${\langle y \rangle}$ is the average value of the partition and $y_i$ is the value
of each data point within the partition. 
In this work, we implement the RF algorithm using sklearn libraries \cite{sklearn}. 
Due to the data's outliers and to decrease the error in the prediction given 
by the ML technique, we re-scale the data by the StandardScaler method. 

As mentioned, we implement the RF to predict the monthly average temperature $\widehat{T}_k$ in
the $k$-th month for each region, by searching a function $f$, such that 
\begin{equation}
\widehat{T}_k = f(g_k, T_{k-{\rm lag}}), \label{eq2}
\end{equation}
where the inputs are the gases $g_k$ and the delayed temperature $T_{k-{\rm lag}}$, 
where lag $=1,2,3,...$ $k$-th month. The delayed temperature corresponds to  
the temperature from the previous months to predict in $k$-month. 
For instance, if the aim is to predict the temperature in April, with  
 lag equal to 3, we use the  
temperatures from January, February, and March, in the same year.

To evaluate the performance of the forecasting, we calculate the absolute error 
($\Delta {\rm} E$), the root mean square error (RMSE), the correlation coefficient ($r$), 
the mean absolute error (MAE) \cite{Willmott2023}, 
the normalized root mean squared error (NRMSE) \cite{Jadon2024}, 
and the refined Willmott's index \cite{Willmott2012}. These metrics are 
discussed in Appendix B.

\section{Time series analysis}\label{time_series_analysis}
Before implementing the techniques to detect breakpoints, we explore some 
properties of the time series. 
Firstly, we focus on $s_t$.
As mentioned, the mathematical form of $s_t$ can be represented by oscillatory functions. 
For example, for South, Southeast, and Brazil the  
adjust is given by  $s_t = b \, {\rm sin}(\omega_1 t + \phi_1)$. 
In this case, our main interest 
is in the frequency, which informs the seasonality period of our time series. 
For South, Southeast, and Brazil, we obtain $\omega_1 = 0.52 \, \pm \, 2.12 \times 10^{-5}$, 
$\omega_1 = 0.52 \, \pm \, 3.54 \times 10^{-5}$, $\omega_1 = 0.52 \, \pm \, 4.18 \times 10^{-5}$, 
respectively. These frequencies imply in a period equal to $T = 12$ months, with 
an error proportional to $10^{-5}$.  For the remaining temperature time series 
(Center-West, Northeast, and North), the mathematical 
form that better describes $s_t$ is a linear combination of two harmonic oscillatory functions, 
such as $s_t = b \, {\rm sin} (\omega_1 t + \phi_1) + c \, {\rm cos}(\omega_2 t + \phi_2)$. 
The first frequency ($\omega_1$) is associated with a period of 12 months, 
while the second ($\omega_2$) with a period of 6 months. In these cases, the 
second component disturbs the seasonality by an amplitude $c$. Furthermore, 
regions with this kind of $s_t$ do not present the four seasons well defined. 
For instance, for Center-West, $\omega_1 = 0.52 \pm 3.6 \times 10^{-5}$, 
$\omega_2 = 1.04 \pm 5.1 \times 10^{-5}$, with $b = 1.49 \pm 0.01$ and $c = 1.06 \pm 0.01$. 
For Northeast, 
we find   
$\omega_1 = 0.52 \pm 6.44 \times 10^{-5}$, 
$\omega_2 = 1.06 \pm 0.001$, with 
$b = 0.75 \pm 0.01$ and $c = -0.02 \pm 0.01$. 
For North, $\omega_1 = 0.52 \pm 10^{-4}$ and $\omega_2 = 1.04 \pm 4.37 \times 10^{-5}$  
with $b = -0.70 \pm 0.003$, and $c = 0.40 \pm 0.004$. These results show that 
the time series South, Southeast, and Brazil exhibit a very well defined 
seasonality of 12 months, while the remaining cases show a component in 
6 months.  It is important to mention 
that the same information is found when we conduct a power-spectra analysis 
for each $y_t$ time series. 

The time series associated with the gases have a seasonal component. 
However, the contribution of $s_t$ into $y_t$ is very low. For example, 
the CH$_4$ time series is in the order of $10^0$ (in ppb units) and the seasonal 
components are proportional to $b = 9 \times 10^{-7} \, \pm \, 6.20 \times 10^{-9}$ 
and  $c = -2.41 \times 10^{-7} \, \pm \, 6.87 \times 10^{-9}$. The same occurs for 
N$_2$O (in ppb), which is in the order of $10^{-1}$ and the mathematical 
form of $s_t$ is in the order of $b = 5.07 \times 10^{-7} \, \pm \, 3.49 \times 10^{-9}$ and 
$c = -1.36 \times 10^{-7} \, \pm \, 3.72 \times 10^{-9}$. 
For CO$_2$ (in ppm), the time series is in the 
order of $10^2$, and its seasonal functions are $b = 3 \times 10^{-3} \, \pm \, 2.60 \times 10^{-5}$ 
and $c = -3 \times 10^{-3} \,\pm \, 2.78 \times 10^{-5}$. Then,  
seasonal components for the gases are practically neglected. 

In terms of trend, the increase is adjusted by $\tau_t = a + b \,t + \Gamma(t)$, 
where $\Gamma(t)$ is a deviation given by an oscillatory function. For the 
temperature time series, the inclination ($b$) is proportional to $10^{-3}$ 
with error proportional to $10^{-5}$. The amplitude of $\Gamma$ is in order 
of 1 $^{\rm o}$C for South and Southeast, while for the remaining time series 
is less than 1. In addition, for the gases the parameter $\Gamma$ is practically null  
and the trend follows almost the same shape as the gas emission (Fig. \ref{fig3}).

By implementing the analysis discussed in Subsection \ref{breakpoints}, 
we identify breakpoints located at 1989-09 for South, 1981-10 for Southeast, 
1993-06 for Center-West, 1997-08 for North, 1986-10 for Northeast and 
2001-10 for Brazil time series, as displayed in Table \ref{tabela0}. Three 
of these time series exhibit breakpoints in 80's, while two in 90's and 
Brazil in 2001-10. These breakpoints are associated with the 
increase in the monthly average temperature and in the trend, as observed 
by the results in Table \ref{tabela0}. In the Table \ref{tabela0}, 
we employ the notation $\langle \cdot \rangle_{\rm b}$ and  $\langle \cdot \rangle_{\rm a}$, 
where the index ``${\rm b}$" means before the breakpoints and ``${\rm a}$" after. 

\begin{figure}[!htb]
\centering
\includegraphics[scale=0.4]{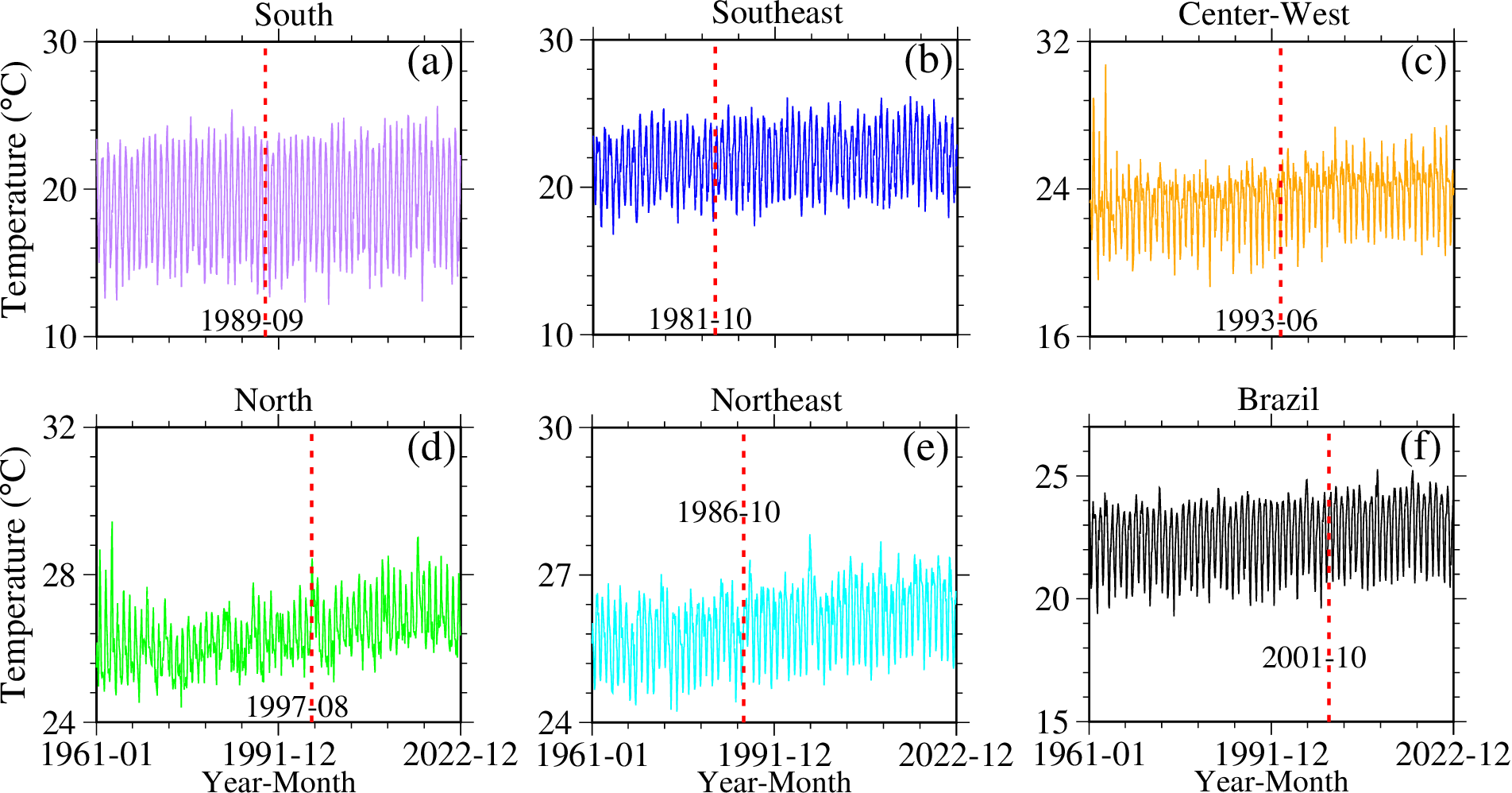}
\caption{Monthly average temperature ($^{\rm o}$C) for the average of region's 
capitals: 
(a) South, 
(b) Southeast, 
(c) Center-West, 
(d) North, and
(e) Northeast. Panel (f) exhibit the the series for the whole Brazil.  
The vertical 
dotted red line marks the breakpoints for each region. 
The values of mean temperature ($\langle T \rangle$) and mean trend ($\langle \tau \rangle$) 
before (and after) the breakpoints are in Table \ref{tabela0}.
}
\label{fig2}
\end{figure}

\begin{figure}[!htb]
\centering
\includegraphics[scale=0.4]{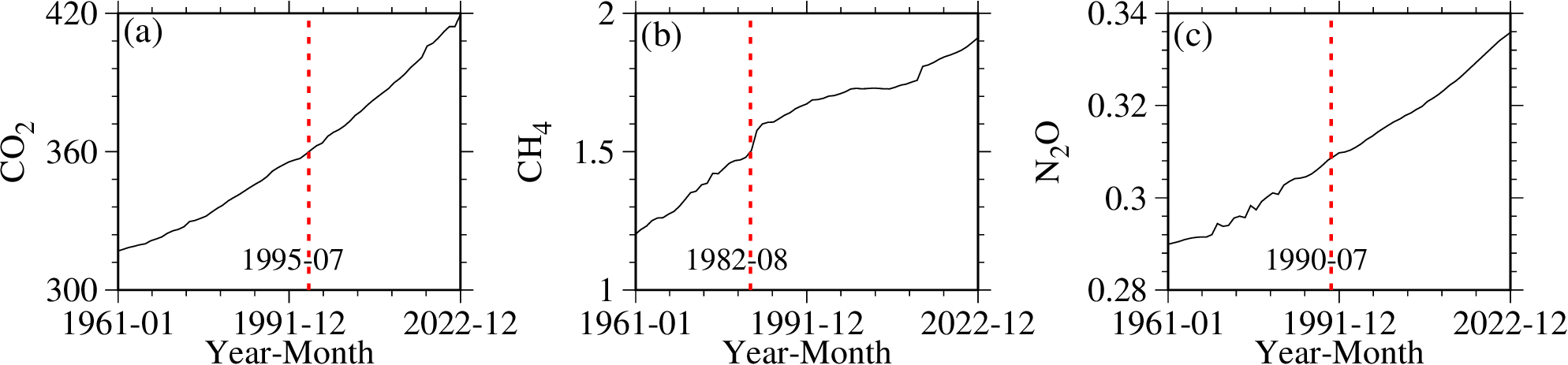}
\caption{Concentration of 
(a) CO$_2$ (ppm), 
(b) CH$_4$ (ppb) and  
(c) N$_2$O (ppb). 
The vertical 
dotted red line marks the breakpoints for each gas. 
The values of mean concentration ($\langle C \rangle$) and mean trend ($\langle \tau \rangle$) 
before (and after) the breakpoints are displayed in Table \ref{tabela01}. }
\label{fig3}
\end{figure}

In addition to the increase in the average monthly temperature, we also 
observe an increase of $\sigma$ and consequently $\sigma^2$, for 
South (Fig. \ref{fig2}(a)), Southeast (Fig. \ref{fig2}(b)) and Northeast (Fig. \ref{fig2}(e)) 
time series. The increase of these quantities after the breaks 
(marked by the vertical red dotted line in Fig. \ref{fig2}) implies an 
increase of the uncertain in relation to the mean value. On the other hand, 
for Center-West (Fig. \ref{fig2}(c)), North (Fig. \ref{fig2}(d)) and 
Brazil (Fig. \ref{fig2}(f)) time series, $\sigma$ decreases after the 
breakpoints. For these regions, the change in $\sigma$ values shows that the 
actual temperature values approximate the mean value 
(i.e., the oscillation decreases the amplitude), which are higher after 
the breakpoints when compared with the data before. 
We verify that the average monthly temperature is 
increasing for all the time series, which is in agreement with 
a global warming scenario. 

\begin{table*}[!htb]
	\centering
	\begin{tabular}{c|c|c|c|c|c}
	\hline
		Time series    & Breakpoints & $\langle T \rangle_{\rm b} \pm \sigma$ & $\langle T \rangle_{\rm a} \pm \sigma$     &$\langle \tau \rangle_{\rm b} \pm \sigma$ & $\langle \tau \rangle_{\rm a}\pm \sigma$	\\ \hline
		South          & 1989-09    & $18.92 \pm 3.18$            & $19.31 \pm 3.26$                &$18.93 \pm 0.38$               & $19.30 \pm 0.40$   	                \\ \hline
		Southeast      & 1981-10    & $21.46 \pm 1.96$            & $22.07 \pm 2.03$                &$21.59 \pm 0.40$               & $22.12 \pm 0.36$  				    \\ \hline
		Center-West    & 1993-06    & $23.11 \pm 1.56$            & $23.95 \pm 1.48$                &$23.12 \pm 0.34$               & $23.96 \pm 0.27$                      \\ \hline
		North          & 1997-08    & $26.10 \pm 0.76$            & $26.80 \pm 0.75$                &$26.10 \pm 0.30$               & $26.79 \pm 0.30$                      \\ \hline
		Northeast      & 1986-10    & $25.71 \pm 0.62$            & $26.21 \pm 0.66$                &$25.72 \pm 0.19$               & $26.21 \pm 0.22$                      \\ \hline
		Brazil         & 2001-10    & $22.43 \pm 1.64$            & $23.00 \pm 1.27$                &$22.43 \pm 0.24$               & $23.00 \pm 0.18$                      \\ \hline
	\end{tabular}
		\caption{Average temperature (trend) before and after the respective 
		breakpoints denoted by  $\langle T \rangle_{\rm b}$ and $\langle T \rangle_{\rm a}$ 
		($\langle \tau \rangle_{\rm b}$ and $\langle \tau \rangle_{\rm a}$). 
		The units are in Celsius ($^{\rm o}$C).}
	\label{tabela0}
\end{table*} 

Now, we employ a similar analysis for the gases, where the breakpoints are
located at  
1995-07, 1982-07 and 1990-07 for CO$_2$, CH$_4$ and N$_2$O, respectively. 
These breakpoints are associated with the increase in the gas emission ($C$)  
and in the trend ($\tau$) of the gases, as provided in Table \ref{tabela01}. 
For these data, $\sigma$ is less than $\sigma$ for $T$, due to the fact that the gas emission follows almost 
a linear increase, as displayed in Figs. \ref{fig3}(a), \ref{fig3}(b) and \ref{fig3}(c), 
for CO$_2$, CH$_4$ and N$_2$O, respectively. This linear increase is verified 
by the fact that the seasonal component of each time series is practically 
null. 
\begin{table*}[!htb]
	\centering
	\begin{tabular}{c|c|c|c|c|c}
	\hline
		Time series    & Breakpoints & $\langle C \rangle_{\rm b} \pm \sigma $ & $\langle C \rangle_{\rm a}\pm \sigma$     &$\langle \tau \rangle_{\rm b}\pm \sigma$ & $\langle \tau \rangle_{\rm a}\pm \sigma$	\\ \hline
		CO$_2$ (ppm)   & 1995-07    & $335.95 \pm 13.16$          & $387.29 \pm 17.00$              &$336.23 \pm 13.05$             & $386.73 \pm 16.54$  	                \\ \hline
		CH$_4$ (ppb)   & 1982-08    & $1.35 \pm 0.09$             & $1.73 \pm 0.08$                 &$1.36 \pm 0.09$                & $1.73 \pm 0.08$  				    \\ \hline
		N$_2$O (ppb)   & 1990-07    & $0.297 \pm 0.005$           & $0.320 \pm 0.007$               &$0.297 \pm 0.005$              & $0.320 \pm 0.007$                      \\ \hline
	\end{tabular}
		\caption{Average concentration (tendency) before and after the respective 
		breakpoints denoted by  $\langle C \rangle_{\rm b}$ and $\langle C \rangle_{\rm a}$ 
		($\langle \tau \rangle_{\rm b}$ and $\langle \tau \rangle_{\rm a}$) 
		for the gas emission. }
	\label{tabela01}
\end{table*} 
\section{Machine Learning prediction}\label{ml_section}
By means of the time series and their breakpoints, we 
prepare our datasets to implement the RF algorithm for the prediction. 
Firstly, we define strategies to include the inputs in 
Eq. (\ref{eq2}). In this work, we study three combinations of inputs: 
$i)$ only gases, i.e., $\widehat{T}_k = f(g_k)$;  $ii)$ only the temperatures 
from the previous three months, i.e.,  $\widehat{T}_k = f(T_{k-3},T_{k-2},T_{k-1})$; 
and $iii)$ a combination of $i)$ and $ii)$, which yields $\widehat{T}_k = f(g_k, T_{k-3},T_{k-2},T_{k-1})$. 
{ As one of our goals is to study how the greenhouses 
affects the forecasting, we propose the previous three inputs.  
Considering these sets of features, we have one that uses only gases, 
other that use only temperature, and a combination of both variables. 
From these features, it is possible to study how the  
greenhouses and temperature influences   
the forecasting by taking them separately (set $i$ and $ii$) or 
together (set $iii$). }

{In addition, the choice of the previous three months is based 
on an autocorrelation analysis of the temperature data. The analysis is 
performed by using a confidence interval of 95\%, and for the majority 
of our dataset, the correlation is high during the first three months. 
It is worth mentioning that other months also show significant correlation, 
particularly due to seasonal effects. From our tests, we choose just the first 
three months due to the fact that this interval is the minimum 
required to enhance the forecasting metrics. }

Defined the inputs, i.e., the features of ML technique, we consider our 
dataset from 1961-04 until 2022-12, where the notation means Year-Month. 
In Section \ref{time_series_analysis}, we start our time 
series in 1961-01. However, for the ML algorithm the beginning  is in 1961-04,  
due to the delayed of three months considered in the temperature time series (strategies $ii$ and $iii$). 
In this way, the time series length is 741 elements, where 80\%   are 
for training (1961-04 until 2010-07, 592 elements) and 20\% are for testing 
(2010-08 until 2022-12, 149 elements).

\subsection{Normal temperature}
By implementing the RF algorithm, we obtain the results shown in Table \ref{tabela1}, 
{ where the predictions for each strategy are quantified by RMSE,
$r$, MAE, NRMSE, 
and $d_r$ (the metrics are discussed in Appendix B). Another possible metric is the mean square relative error. 
However, by perfoming some calculations our results show a similar result with 
NRMSE, for this reason, we keep just the NRMSE.}
 These quantities are calculated in the testing 
range, i.e., they are metric for the prediction precision in a horizon forecasting 
of 149 months. 
We consider the whole time series, 
including the breakpoints. { By considering only gases as input in our 
ML technique (strategy $i$), our results reveal that is possible to forecast the temperature 
with reasonable metrics. Inspecting each metric, we observe 
a good agreement between the simulated and the real data. For example, 
the MAE suggests an error less than 1.243, while the NRMSE $<0.1$ and 
$d_r \geq 0.668$.}
An improvement in terms of $r$, and the other metrics, for the forecasting is obtained when we include the three previous 
months as input (strategy $ii$). In this case, $r\geq0.8$  and RMSE $\leq1.18$ for all regions. 
{ Now, we achieve MAE $\leq 0.927$ and NRMSE $\leq 0.062$. } 
In terms of correlation, 
the best predicted time series is Brazil, where $r=0.944$. 
{  By combining the three months of delay with the gases (strategy $iii$),  
a minor improvement in the forecasting is obtained for some regions, as observed 
by the metrics in Table \ref{tabela1}. }  
Furthermore, we verify other combinations, including $f(g_k,g_{k-\rm{lag}})$, $f(g_k, T_{k-1})$, $f(g_k, T_{k-2})$, 
$f(g_k, T_{k-3})$, $f(T_{k-1})$, $f(T_{k-2})$, or $f(T_{k-3})$, however 
they generate worst results. All the hyperparameters used in RF are described in the 
Table \ref{tabela1} caption. 
\begin{table*}[!htb]
	\centering
	\begin{tabular}{c|ccccc}
	\hline
		Input          &       &       &$f(g_k)$  & &                      	                          \\ \hline
		               & RMSE  &$r$    & MAE       & NRMSE & $d_r$                \\ \hline
		South          & 1.578 &0.867  & 1.243     & 0.083 & 0.768							\\ \hline
		Southeast      & 1.014 &0.863  & 0.809     & 0.046 & 0.763							\\ \hline
		Center-West    & 1.090 &0.691  & 0.809     & 0.047 & 0.668							\\ \hline
		North          & 0.417 &0.840  & 0.297     & 0.015 & 0.766							 \\ \hline
		Northeast      & 0.332 &0.883  & 0.279     & 0.012 & 0.749							 \\ \hline
		Brazil         & 0.620 &0.880  & 0.501     & 0.027 & 0.771							\\ \hline
		\hline
		Input          &       &       &$f(T_{k-3},T_{k-2},T_{k-1})$  & &                      	                          \\ \hline
		               & RMSE  &$r$    & MAE    & NRMSE & $d_r$                \\ \hline
		South          & 1.181 &0.940  & 0.927  & 0.062 & 0.845  				\\ \hline
		Southeast      & 0.963 &0.885  & 0.792  & 0.044 & 0.782  				\\ \hline
		Center-West    & 1.123 &0.800  & 0.826  & 0.047 & 0.697							\\ \hline
		North          & 0.509 &0.820  & 0.393  & 0.019 & 0.712							 \\ \hline
		Northeast      & 0.342 &0.880  & 0.263  & 0.013 & 0.783							 \\ \hline
		Brazil         & 0.462 &0.943  & 0.358  & 0.020 & 0.849							\\ \hline
		\hline
		Input          &       &       &$f(g_k, T_{k-3},T_{k-2},T_{k-1})$  & &                      	                          \\ \hline
		               & RMSE  &$r$    & MAE    & NRMSE & $d_r$                \\ \hline
		South          & 1.156 &0.942  & 0.908  & 0.060 & 0.848							\\ \hline
		Southeast      & 0.952 &0.890  & 0.761  & 0.043 & 0.790							\\ \hline
		Center-West    & 1.147 &0.781  & 0.807  & 0.049 & 0.704							\\ \hline
		North          & 0.500 &0.825  & 0.380  & 0.018 & 0.721							 \\ \hline
		Northeast      & 0.323 &0.893  & 0.253  & 0.012 & 0.791							 \\ \hline
		Brazil         & 0.454 &0.944  & 0.363  & 0.020 & 0.847							\\ \hline
	\end{tabular}
		\caption{Root Mean Square Error (RMSE), correlation coefficient ($r$), 
		mean absolute error (MAE),  
		normalized root mean square error (NRMSE), and Willmott's index ($d_r$)  
		for the temperature forecasting from 2010-08 until 2022-12, by considering 
		as inputs: only gases ($f(g_k)$), temperature of the last three months 
		($f(T_{k-3},T_{k-2},T_{k-1})$), temperature of the last three months and 
		gases ($f(g_k, T_{k-3},T_{k-2},T_{k-1})$). 
		For $f(g_k)$ and $f(T_{k-3},T_{k-2},T_{k-1})$ the hyperparameters 
		(criteria to choose partitions, number of trees, maximum features, and depth)  
		for South are: Absolute error, 300, 0.2, 30; 
		for Southeast are: Absolute error, 300, 0.2, 30;
		for Center-West are: Friedmann MSE, 300, 0.2, 30; 
		for North are: Absolute error, 200, 0.5, 30; 
		for Northeast are: Absolute error, 300, 0.6, 30; 
		for Brazil are: Absolute error, 200, 0.6, 30.
		For $f(g_k, T_{k-3},T_{k-2},T_{k-1})$ the maximum features changes 
		to 0.7 in South; to 0.5 in Southeast, Center-West, and North.}
	\label{tabela1}
\end{table*}  

\begin{figure}[!htbp]
\centering
\includegraphics[scale=0.45]{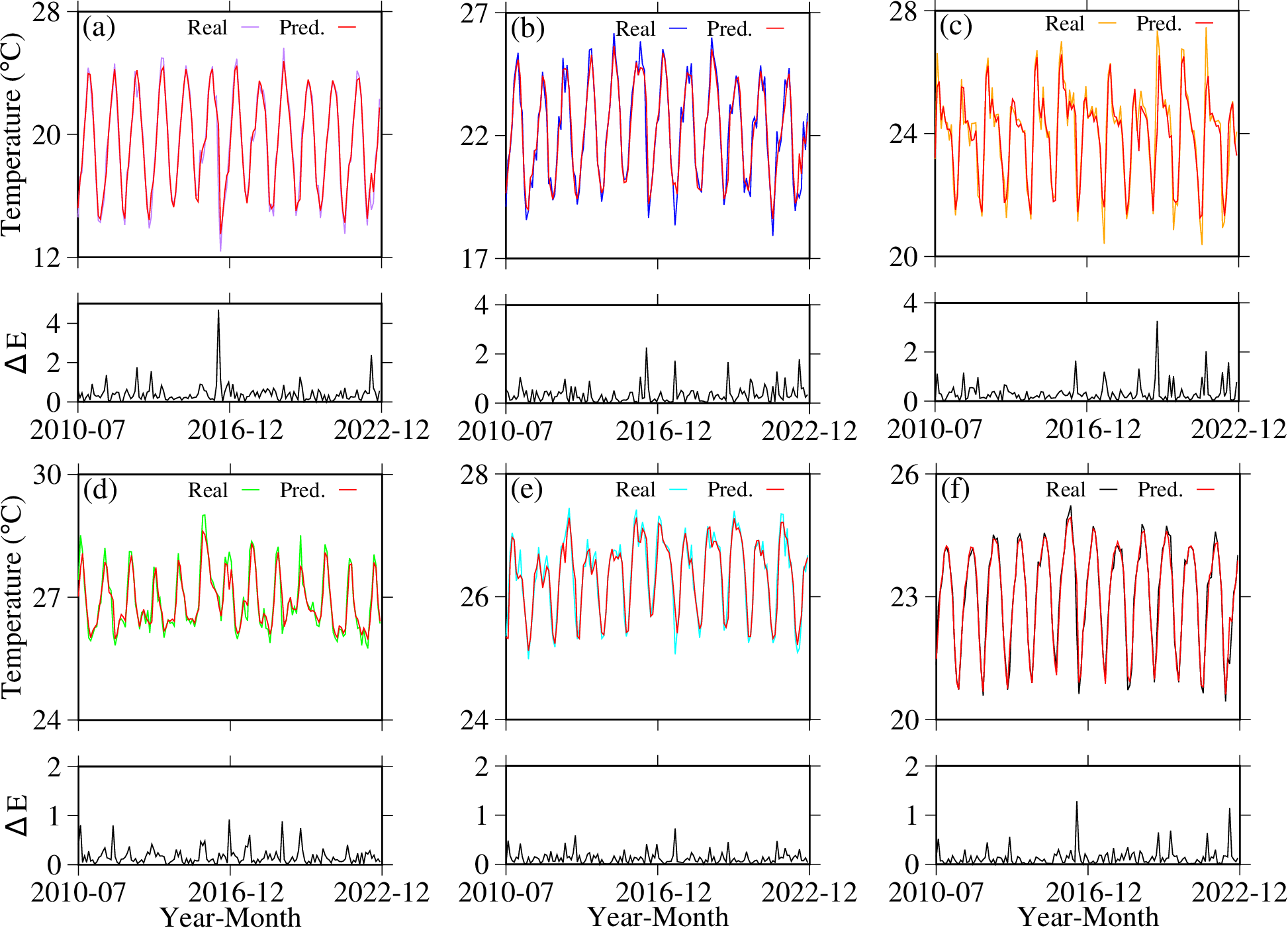}
\caption{Implementation of ML algorithm to predict (red lines) the 
temperatures of 
(a) South, 
(b) Southeast, 
(c) Center-West, 
(d) North, 
(e) Northeast, and 
(f) Brazil. The sub-floats display the absolute error between real temperature 
and predicted. The results are for the method $(g_k, T_{k-3},T_{k-2},T_{k-1})$.}
\label{fig4}
\end{figure}

Considering the prediction based on strategy $iii$, we compute the predicted 
temperatures, as displayed in Fig. \ref{fig4} by red line for 
(a) South, 
(b) Southeast, 
(c) Center-West, 
(d) North, 
(e) Northeast, 
and (f) Brazil. In each panel, the sub-floats show $\Delta \rm{E}$ 
(Eq. (\ref{eq_erro})), which increases after 2016-11. 
Before this point, $\Delta \rm{E}$ is  low, e.g., 
for the results  in Figs. \ref{fig4}(a), (b) and (c), 
we get $\Delta \rm{E} < 3.5$ $^{\rm o}$C, while for Figs. \ref{fig4}(d), 
(e) and (f), we find $\Delta \rm{E} < 1.6$ $^{\rm o}$C. 
The maximum values of error are almost reached just in 
some points of the time series. For the most part of the simulation, $\Delta \rm{E}$ 
is lower than these values. 
Our analyses enable us to affirm that 
the strategy $iii$ produces the most precise forecasting of the temperature. 

\begin{figure}[!htbp]
\centering
\includegraphics[scale=0.4]{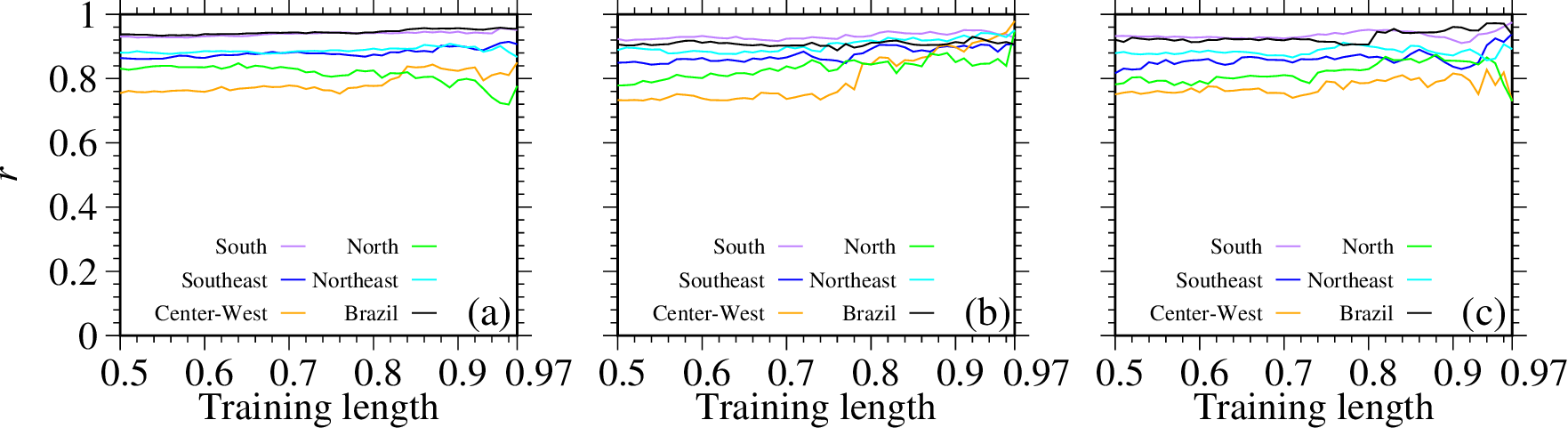}
\caption{$r$ as function of training length for 
South (purple line), Southeast 
(blue line), Center-West (orange line), North (green line), Northeast 
(cyan line), and Brazil (black line). 
The panel (a) exhibits the whole time series. 
The panel (b) displays the test by splitting the time series before the breakpoints, 
while the panel (c) shows after it.}
\label{fig5}
\end{figure}

The results in Fig. \ref{fig4} are obtained using 80\% of the 
time series length to training the RF algorithm. However, for some regions,  
$r$ is improved by considering different training length.  
One of these regions is Center-West, where $r$ increases from 0.781 to  
0.840 for a training length greater than $0.8$ (orange curve in Fig. \ref{fig5}(a)).
Other regions, such as South, Southeast, and Brazil have just a little 
increase in $r$ for long training length, as displayed by the purple, blue,  
and black curves in Fig. \ref{fig4}. Considering a long training length 
for Southeast ($>0.935$), the results show a decrease in $r$, while the 
other time series increase or 
conserve the $r$ value in these range of training. One particular case is 
North, where $r$ decreases in the training length $\in (0.8, 0.9525)$. 
All in all, we observe that the training length affects more 
the prediction for some time series than others.  

As previously observed, our time series exhibit some breakpoints. To investigate 
if these points affect the ML precision, we split the time series before 
and after the breakpoints. To split, we are taking into account the breakpoints associated 
with the gases. In this way, the time series before the 
breakpoints are: 
South from 1961-04 until 1982-07; 
Southeast from 1961-04 until 1981-08; 
Center-West from 1961-04 until 1982-07; 
North from 1961-04 until 1982-07; 
Northeast from 1961-04 until 1982-07; 
Brazil from 1961-04 until 1982-07. 
And after: 
South from 1995-08 until 2022-12; 
Southeast from 1995-08 to 2022-12; 
Center-West from 1995-08 until 2022-12; 
North from 1997-07 until 2022-12; 
Northeast from 1995-08 until 2022-12; 
Brazil from 2001-11 until 2022-12. 

Considering the time series before the breakpoints and employing the ML 
technique, we compute $r$ as a function of the training length for each dataset, 
obtaining the result in Fig. \ref{fig5}(b). 
For all considered cases, $r$ increases for a training length superior 
than 0.78. Therefore, $r$ shows a significant increase, being 
$r>0.84$ for a training length $\leq 0.96$.
One particular case is Center-West, where $r$ goes from $0.85$ (length training $=0.8$) 
to $r = 0.978$ (length training $=0.97$). The time 
series have a clear pattern for time inferior to 1982-07 (Fig. \ref{fig2}(c)), 
which is well learned by RF. In addition, the testing range 
has a smooth behaviour, which is easier for the ML to predict.  

Now, we consider the dataset after the breakpoints, and find the results  
shown in Fig. \ref{fig5}(c). For all the cases, $r$ exhibits  
a significant oscillation after training length $>0.8$. South (purple curve), 
Southeast (blue curve), and Northeast (cyan curve) have a similar behaviour 
for a training length $>0.85$. For these time series, $r$ decreases for  
training lengths in $\in (0.88, 0.93)$, and after that increases. 
For training length $>0.96$, South and Southeast present a high value 
of $r$, i.e., superior than 0.95 and 0.92, respectively. However, Northeast 
shows an opposite change for the training length $=0.96$. In this value,  
$r =$ 0.904 and after that,  for the training length $=0.97$, becomes 0.890.  
The remaining time series (Center-West [orange], North [green] and 
Brazil [black]) decrease in $r$ for a training length $>0.94$. 
For Center-West these changes go from $r=0.824$ until 0.728. 
For North this change is higher, going from $r=0.865$ to 0.728. 
A small change occurs for whole Brazil, which goes from $r=0.971$ to 0.937. 

\subsection{Anomalous temperature}
\begin{figure}[!htb]
\centering
\includegraphics[scale=0.4]{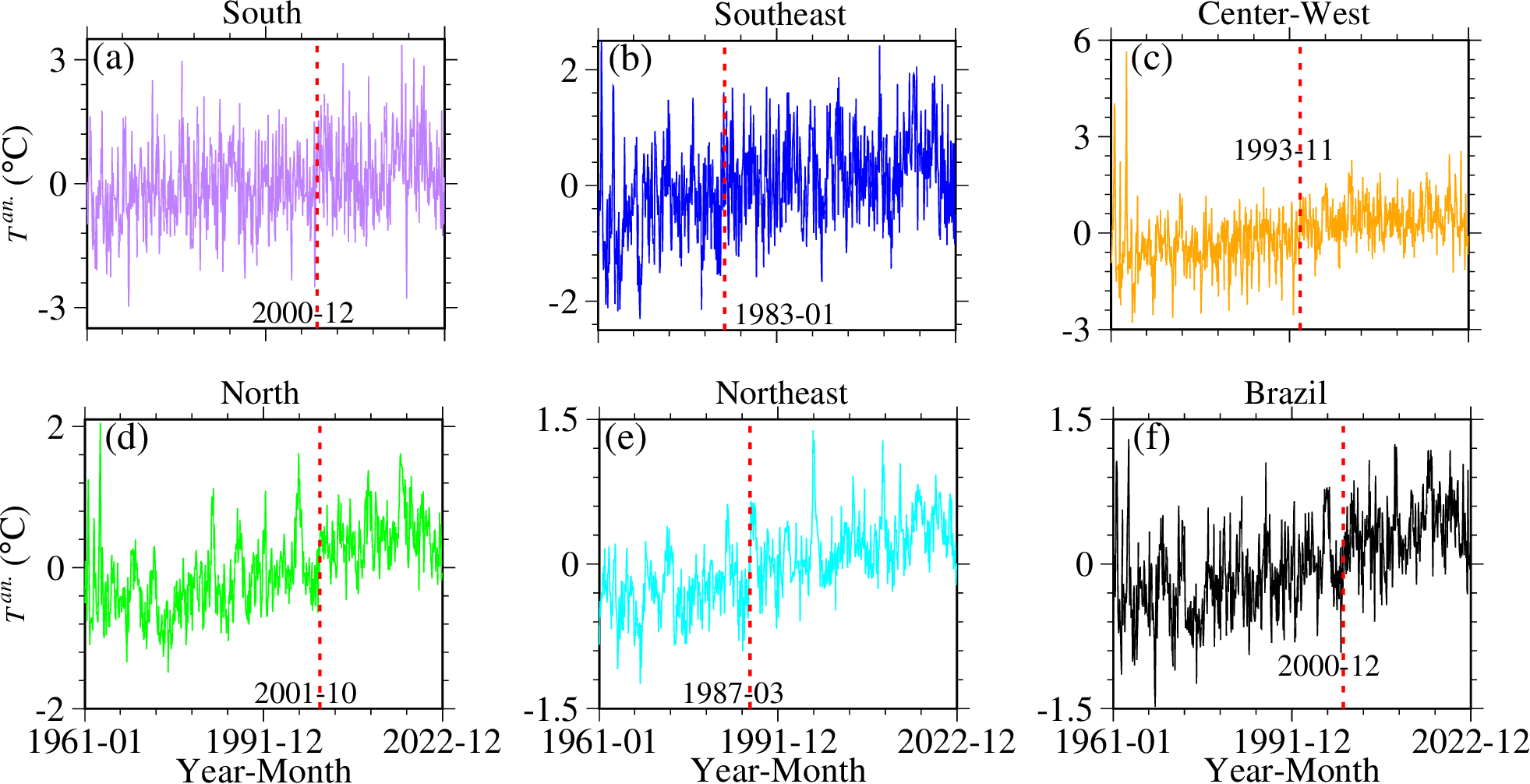}
\caption{Monthly anomalous temperature ($T^{an.}$ $^{\rm o}$C) for the average of region's 
capitals: 
(a) South, 
(b) Southeast, 
(c) Center-West, 
(d) North, and
(e) Northeast. The panel (f) exhibits  the series for the whole Brazil.  
The vertical 
dotted red line marks the breakpoints for each region. 
}
\label{fig6}
\end{figure}

Considering the algorithm to predict the temperatures, we remove 
the annual component from each time series in order to verify the 
improvement (or not) in the forecasting. The new dataset without the annual 
component, namely anomalous, is calculated from the following transformation 
\begin{equation}
T^{an.}_{m} (l) = T_{m} (l) - \langle T_{m} (l)\rangle,
\label{an_eq}
\end{equation}
where $T^{an.}$ is the anomalous temperature, $m$ is the month, $l$ is the 
year, $\langle T_{m} (l)\rangle$ is the average of temperature in  
months overall all the years. Applying Eq. (\ref{an_eq}) in the original 
dataset, we get the anomalous temperatures, as exhibit in Figs. \ref{fig6}(a)-(f) for  
South, 
Southeast, 
Center-West, 
North, 
Northeast, and 
Brazil, respectively.

Firstly, we investigate the breakpoints associated with the new dataset. 
Each breakpoint is highlighted by the vertical red dotted line in Fig. \ref{fig6}. 
The breakpoints associated with South, Southeast, Center-West, North, and Northeast, 
occur late when compared with normal data. These breakpoints are now located 
at 2000-12, 1983-01, 1993-11, 2001-10 and 1987-03, respectively. The 
only exception occurs for Brazil, where the breakpoints  is in 2000-12.  
Previously, it was in 2001-10. Additionally, by removing the annual component 
the new time series oscillate rather irregularly when compared with normal data. 

Due the fact that this new dataset is more noisy when compared with the 
normal dataset, the prediction exhibits {higher RMSE, MAE, 
NRMSE and lower $r$ and $d_r$ (Table \ref{tabela2}) compared with normal 
data. In this case, the metrics indicate that certain regions are better 
described by a given set of features, while others are better described 
by another configuration. However, inspecting the metrics, the predicted 
data presents better metrics through the first strategy. We observe that 
the data from the Northeast is better described by the three sets of 
features. Considering this time series, the best metrics are achieved 
using just the gases as input.} 
For the RF algorithm, again we use 80\% of the time series length for 
training and 20\% for testing. The hyperparameters for each combination of 
input are described in Table \ref{tabela2}.
\begin{table*}[!htb]
	\centering
	\begin{tabular}{c|ccccc}
	\hline
		Input          &       &      &$f(g_k)$  & &   					         \\ \hline
		               & RMSE  &$r$   & MAE    & NRMSE & $d_r$         \\ \hline
		South          & 0.866 &0.400 & 0.697  & 0.169 & 0.513		\\ \hline
		Southeast      & 0.675 &0.555 &	0.547  & 0.180 &	0.572	\\ \hline
		Center-West    & 0.800 &0.540 & 0.585  & 0.160 &	0.605	\\ \hline
		North          & 0.293 &0.847 & 0.222  & 0.115 &	0.748	\\ \hline
		Northeast      & 0.197 &0.880 & 0.152  & 0.090 &	0.763	\\ \hline
		Brazil         & 0.322 &0.717 & 0.246  & 0.138 &	0.669\\ \hline
		\hline
		Input          &       &       &$f(T^{an.}_{k-3},T^{an.}_{k-2},T^{an.}_{k-1})$  & &   					         \\ \hline
		               & RMSE  &$r$    & MAE    & NRMSE & $d_r$            \\ \hline
		South          & 0.926 &0.189  & 0.733  & 0.205 & 0.500 				\\ \hline
		Southeast      & 0.726 &0.423  & 0.605  & 0.187 & 0.534 			   \\ \hline
		Center-West    & 0.897 &0.500  & 0.666  & 0.110 & 0.547			 \\ \hline
		North          & 0.367 &0.776  & 0.284  & 0.111 & 0.687			  \\ \hline
		Northeast      & 0.231 &0.823  & 0.171  & 0.088 & 0.717		   \\ \hline
		Brazil         & 0.355 &0.630  & 0.292  & 0.154 & 0.597			  \\ \hline
		\hline
		Input          &       &       &$f(g_k, T^{an.}_{k-3},T^{an.}_{k-2},T^{an.}_{k-1})$  & &   					         \\ \hline
		               & RMSE  &$r$    & MAE    & NRMSE & $d_r$          \\ \hline
		South          & 0.880 &0.302  & 0.690  & 0.195 & 0.529					 \\ \hline
		Southeast      & 0.700 &0.480  & 0.571  & 0.181 & 0.560				 \\ \hline
		Center-West    & 0.888 &0.513  & 0.621  & 0.110 & 0.579				 \\ \hline
		North          & 0.374 &0.764  & 0.281  & 0.112 & 0.691				 \\ \hline
		Northeast      & 0.227 &0.830  & 0.170  & 0.086 & 0.718					 \\ \hline
		Brazil         & 0.348 &0.646  & 0.272  & 0.150 & 0.625				 \\ \hline
	\end{tabular}
		\caption{Root Mean Square Error (RMSE), correlation coefficient ($r$), 
		mean absolute error (MAE), 
		normalized root mean square error (NRMSE), and Willmott's index ($d_r$)
		for the anomalous temperature forecasting from 2010-08 until 2022-12, by considering 
		as inputs: only gases ($f(g_k)$), temperature of the last three months 
		($f(T_{k-3},T_{k-2},T_{k-1})$), temperature of the last three months and 
		gases ($f(g_k, T_{k-3},T_{k-2},T_{k-1})$). 
		For $f(g_k)$ the hyperparameters 
		(criteria to choose partitions, number of trees, maximum features, and depth)  
		for South are: Absolute error, 200, 0.2, 30; 
		for Southeast are: Absolute error, 500, 0.2, 30;
		for Center-West are: Absolute error, 600, 0.2, 30; 
		for North, Northeast, and Brazil are: Absolute error, 500, 0.2, 30.
		For $f(T_{k-3},T_{k-2},T_{k-1})$ the hyperparameters for 
		South are: Friedman MSE, 500, 0.2, 50;
		for Southeast are: Absolute error, 200, 0.2, 30;
		for Center-West, North and Northeast are: Absolute error, 200, 0.2, 30; 
		for Brazil are: Friedman MSE, 200, 0.2, 30.
		For $f(g_k, T_{k-3},T_{k-2},T_{k-1})$ the hyperparameters 
		for South are: Absolute error, 200, 0.2, 30; 
		for Southeast are: Absolute error, 500, 0.2, 30;
		for Center-West: Friedman MSE, 400, 0.2, 30;
		for North and Brazil: Friedman MSE, 500, 0.2, 30;
		for Northeast: Friedman MSE, 800, 0.2, 30.}
	\label{tabela2}
\end{table*}

\begin{figure}[!htbp]
\centering
\includegraphics[scale=0.4]{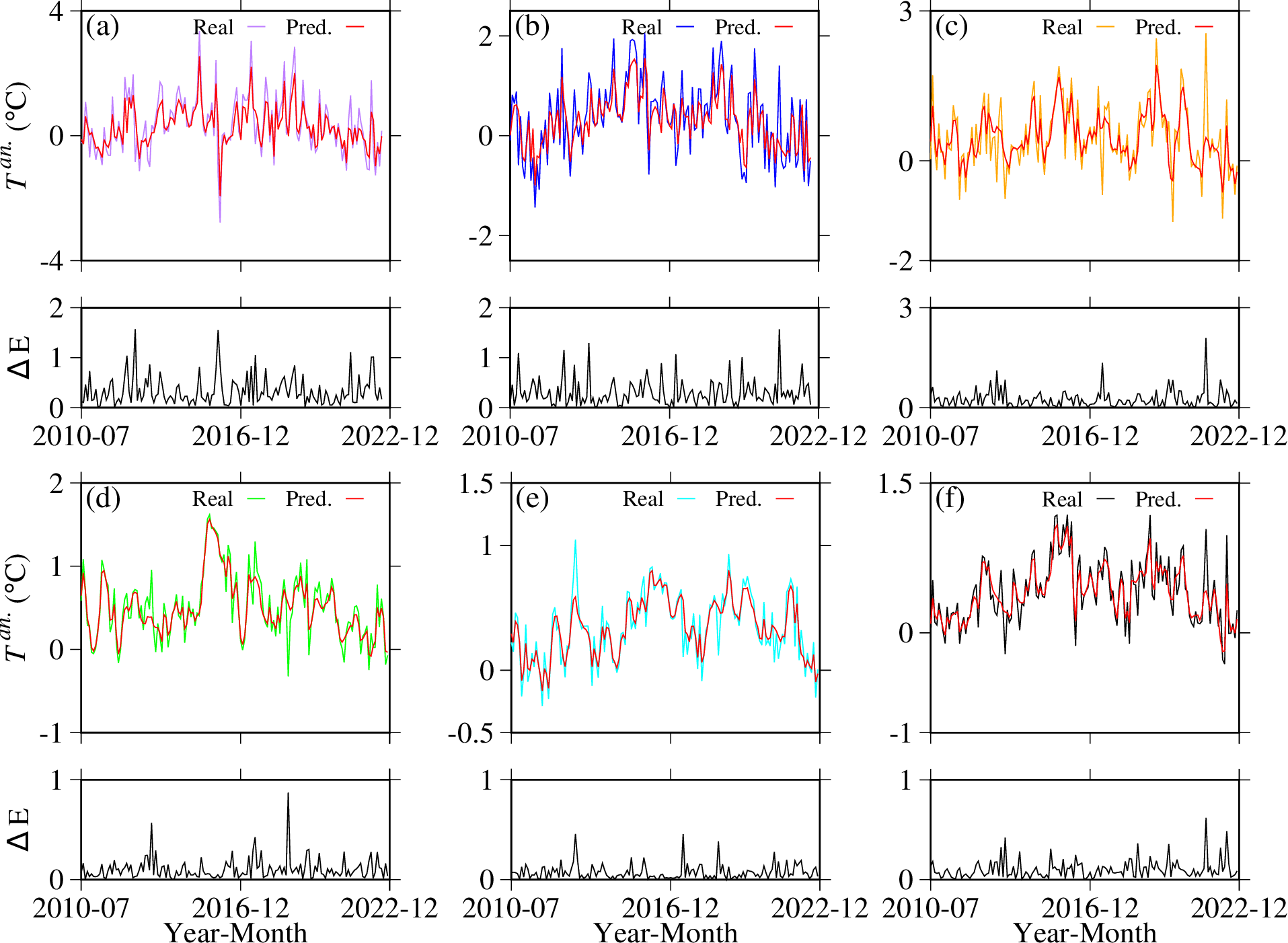}
\caption{Implementation of ML algorithm to predict (red lines) the 
anomalous temperatures of 
(a) South, 
(b) Southeast, 
(c) Center-West, 
(d) North, 
(e) Northeast, and 
(f) Brazil. The sub-floats display the absolute error between real temperature 
and predicted. The results are obtained for strategy $(g_i)$.}
\label{fig7}
\end{figure}

From Table \ref{tabela2}, most part of the data are better 
described through the strategy $i$. Following this strategy, Fig. \ref{fig7} shows the forecasting 
for 
(a) South, 
(b) Southeast, 
(c) Center-West, 
(d) North, 
(e) Northeast, 
and (f)  Brazil. Besides $r$ is not 
high for the considered cases, we see that the ML technique is able to 
reproduce the shape of the curve for all the regions. 
One of the reasons to lose the correlation is because the time series have a higher noise, 
which are not well learned by the RF.    

\begin{figure}[!htbp]
\centering
\includegraphics[scale=0.4]{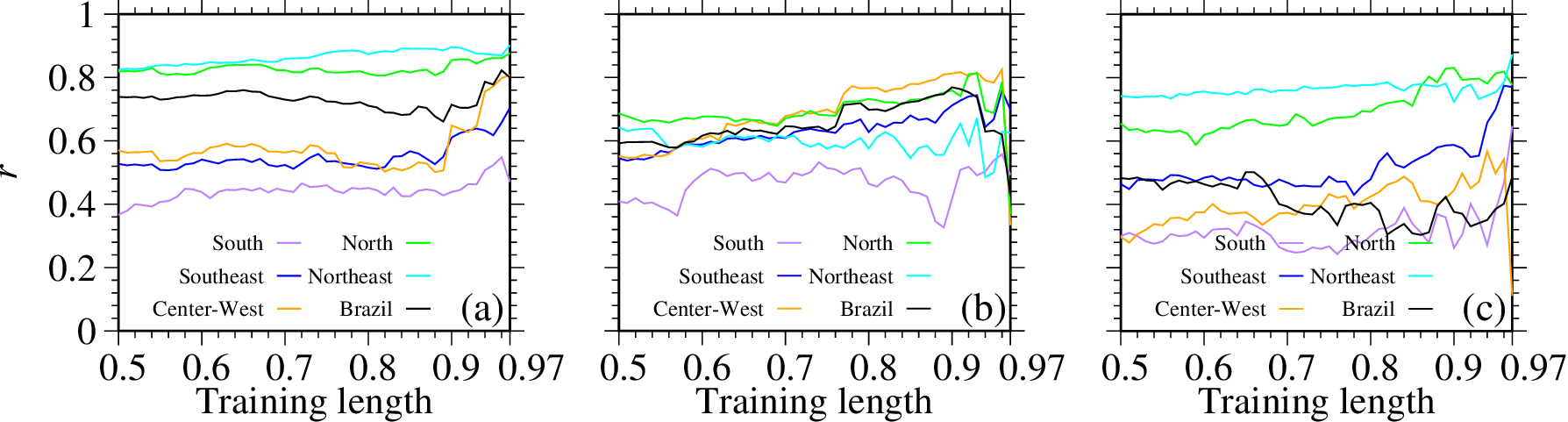}
\caption{$r$ as function of training length considering anomalous temperature data for 
South (purple line), Southeast 
(blue line), Center-West (orange line), North (green line), Northeast 
(cyan line), and Brazil (black line). 
The panel (a) shows the whole time series. 
The panel (b) displays the test by splitting the time series before the breakpoints, 
while the panel (c) shows after it.}
\label{fig8}
\end{figure}

The value of $r$  increases when more points are used in the training process  
(Fig. \ref{fig8}(a)). 
For example, in the training range $(0.88, 0.97)$, $r$ increases for all time series, 
in particular for Center-West, which  
goes from $r=0.5$ to $r=0.8$. For a training length $>0.9$ the only time 
series that does not reach $r=0.6$ is South, that always stays below $r=0.56$. 
In Fig. \ref{fig7}(a), the anomalous South dataset does not 
present any pattern and looks more noisier when compared with the others. 
Due to this fact, the prediction lost precision. 
 
Using the strategy $i$,  we verify 
the algorithm performance before and after the breakpoints. Firstly, 
we split the time series before and after it. 
For South, the data is divided from 1961-01 to 1982-08 and from 2000-12 
to 2022-12. 
For Southeast and Center-West, the time series is split from 1961-01 to 1982-08 and from 
1995-07 to 2022-12. 
For North, we use the period from 1961-01 up to 1982-08 and from 2001-10 until 2022-12. 
For Northeast, the data is divided from 1961-01 until 1982-08, and from 1995-07 
up to 2022-12. For Brazil time series, we split from 1961-01 up to 1982-08 and from 2000-12 
until 2022-12. This setup allows us to explore the algorithm performance. 

In general, training the algorithm before (Fig. \ref{fig8}(b)) and after 
(Fig. \ref{fig8}(c)) the breakpoints generate worst results when compared 
with the whole time series (Fig. \ref{fig8}(a)). In this case, these breakpoints  
do not add any error in the training process. This is observed 
for a wide range of training length. One exception occurs 
for Center-West, where the $r$ value, orange line in Fig. \ref{fig8}(b), 
increases when compared with the 
orange curve from Fig. \ref{fig8}(a).   
\section{Conclusion}\label{conclusao}
In this work, we employ a breakpoints analysis and prediction of  
temperature time series. The prediction is based on Random Forest, which 
is a commonly used Machine Learning (ML) algorithm. 
It is employed to predict the temperature in the $k$-th month, 
by considering the greenhouse emissions 
(CO$_2$, CH$_4$, and N$_2$O) and the previous three months' temperature.  
As a dataset, we use the monthly average temperature from the 27 states 
capitals of Brazil, recorded from January 1961 to December 2022 and 
from the whole Brazil. The gas emissions are monthly recorded in the same time window. 
We group the 
capital time series according to the geographical region 
(South, Southeast, Center-West, North, and Northeast). In this way, 
we analyse 6 temperature time series. To verify 
the predictive ability of the RF model, we validated it in the range 2010-07 
until 2022-12, totaling a horizon forecast of 149 months.

Before the forecasting, we compute the  breakpoints for 
all time series, getting 1989-09, 1981-10, 1993-06, 
1997-08, 1986-10 and 2001-10 for South, Southeast, Center-West, North, 
Northeast, and Brazil. For the breakpoints, we verify that they are 
associated with an increase in the mean temperature. For the 
gases, the breakpoints are in 1995-07, 1982-08 and 1990-07 for CO$_2$, CH$_4$, and N$_2$O, 
respectively. These breaks are associated with a significant increase 
in emissions. 

We define three 
different strategies of features in the RF algorithm: 
$i)$ global emission of CO$_2$, CH$_4$, and N$_2$O; $ii)$ the temperatures from the previous 
three months; and $iii)$ a combination of $ii$ and $iii$. 
To verify the predictive ability of the 
algorithm, we employ the Root Mean Square Error (RMSE), 
{ the correlation coefficient ($r$), 
the mean absolute error (MAE), the normalized root mean squared error (NRMSE), 
and the refined Willmott's index ($d_r$). 
For $i$, the best predicted data is for Northeast, 
while the worst is for Center-West. 
Considering $ii$ and $iii$, the best and the worst 
results are related to Brazil and Center-West, respectively. 
The metrics concerning these results are displayed in Table \ref{tabela1}}.  
Based on these results, 
it is not possible to select the best set of features to forecast all 
the regions. On the other hand, the outcomes exhibit that the temperature 
of each region is better forecasted for a particular set of features. 
{ The considered gases emissions are global indicators; 
despite that, they can be used to forecast the local temperature, as shown 
by our simulations. However, the precision of forecasting is enhanced, 
for most cases, when local indicators are included, i.e., previous temperature data.}

In addition to the previous results, we verify the influence of the annual 
component in the time series. To do that, we generate a new dataset without 
the annual component, known as anomalous data. 
The breakpoints change for all the considered data, becoming late for 
some cases. For this dataset,  RF does not show a good predictive ability 
for this dataset compared to normal data. 
{ For example, for the anomalous data the best described is Northeast, while 
the worst is South, for the three sets of inputs, with the metrics given by 
Table \ref{tabela2}.}

Our work shows that the temperature variations can be predicted 
by using gas emissions, temperatures delayed by three months, and 
a combination of both as input in the ML technique. The best strategy depends 
on the region to be described. Furthermore,  we show the existence of breakpoints in 
temperature and gas data, which show us that the system's behavior  
changed, highlighted by the increases in temperature in a general way. 
{ One limitation of this work is the use of just one method, 
however, future works should be conducted by using other algorithms, 
e.g., XGBoost or Reservoir Computing.}
The results presented in this paper also open new problems. For example, 
our techniques can be used to predict the increase of gases, and then 
scenarios for future temperature increase and control by reducing gas 
emissions can be studied.  
\section*{Acknowledgements}
This work was possible with partial financial support from the following 
Brazilian government agencies: CNPq, CAPES, Funda\-\c c\~ao A\-rauc\'aria 
and S\~ao Paulo Research Fo\-undation (FAPESP 2018/03211-6, 2022/13761-9, 2024/14478-4, 2024/05700-5). 
E. C. G. received partial financial support from Coordena\c c\~ao de
Aperfei\c coamento de Pessoal de N\'ivel Superior - Brasil (CAPES) - Finance
Code 88881.846051/2023-01, and FAPESP under grant 2025/02318-5. 
R. L. V. thanks the financial support from the Brazilian Federal 
Agencies (CNPq) under Grant Nos. 403120/2021-7, 301019/2019-3. 
I. L. C. thanks the financial support from CNPq under grant 302665/2017-0. 
A. M. B. 
thanks the financial support from MCTI/CNPq/BRICS-STI. 
We thank 105 Group Science (www.105groupscience.com).

\section*{Data availability}
The datasets generated during and/or analyzed during the current study are available 
on GitHub \cite{sidneygithub} and also can be requested from the corresponding author.


\newpage
\section{Appendix A}\label{apendice}
\subsection{South Capitals}\label{sul_capitais}
\begin{figure}[htbp]
\centering
\includegraphics[scale=.4]{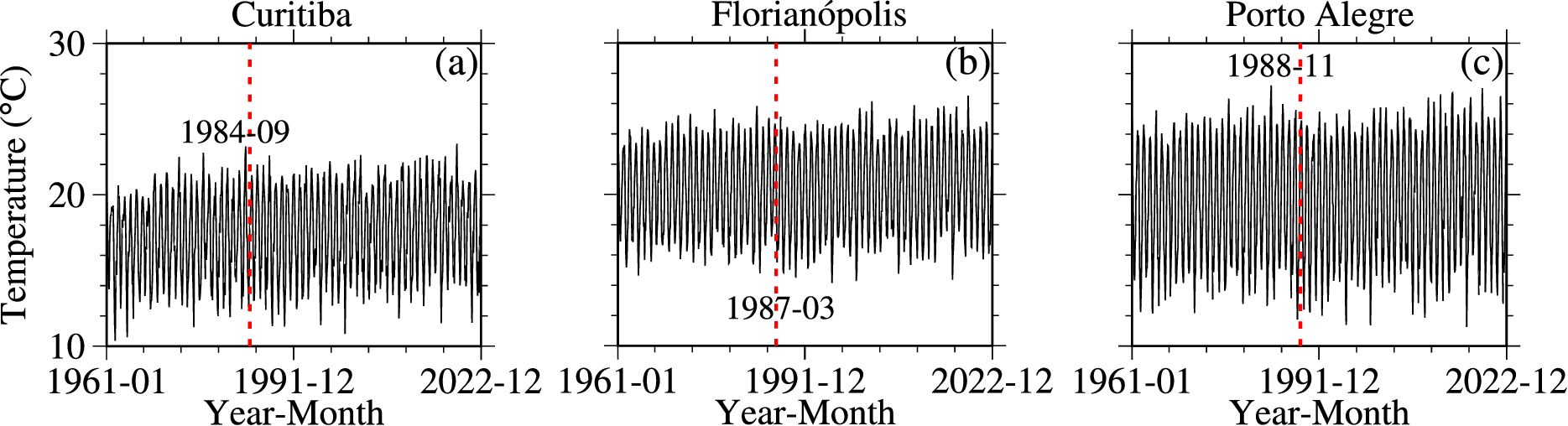}
\caption{Monthly average temperature ($^{\rm o}$C) for the South capitals: 
(a) Curitiba, (b) Florian\'opolis, and (c) Porto Alegre. The vertical 
dotted red line marks the breakpoint for each city, with the respective 
mean temperature ($\langle T \rangle$) and trend ($\langle \tau \rangle$) 
before and after the breakpoint.}
\label{fig9}
\end{figure}

Figure \ref{fig9} displays the temperature ($^{\rm o}$C) time series associated 
with the South Capitals: (a) Curitiba, (b) Florian\'opolis, and (c) Porto 
Alegre, from 1961-01 up to 2022-12. Employing the ADF test, our results 
show that the three time series are stationary. In addition, from our 
simulations, the breakpoints for Curitiba, Florian\'opolis and Porto Alegre 
occur in 1984-09, 1987-03, and 1988-11, respectively. The vertical dotted 
red line marks the point of breakpoints occurrence. Note that all the 
breakpoints occur in 80's decade. Considering the data from Curitiba (panel (a), 
the average temperature (in degree Celsius unit) before the breakpoints is 
$\langle T \rangle = 17.07 \pm 2.80$ and the average trend in this range 
is $\langle \tau \rangle = 17.08 \pm 0.45$. After the breakpoints, the  
values are updated to  $\langle T \rangle = 17.65 \pm 2.89$ and 
$\langle \tau \rangle = 17.66 \pm 0.41$. Besides we observe an increase 
in the average temperature, it is not possible to affirm that a significant  
change occurs due the fact the the standard deviation is considerable high. However, 
we observe   a change in the trend of this time series. Panel (b) 
exhibits a breakpoints in 1897-03 for Florian\'opolis time series. Before 
1897-03 the statistical values are $\langle T \rangle = 20.27 \pm 2.87$ 
and $\langle \tau \rangle = 17.13 \pm 0.47$. After 1897-03 the values become 
$\langle T \rangle = 20.43 \pm 3.05$ and 
$\langle \tau \rangle = 17.65 \pm 0.42$. Again, the breakpoints captures  
the change in the trend of the time series. 
Porto Alegre (panel (c)), which our test shows a breakpoints in 
1988-11. This time series present 
$\langle T \rangle = 19.44 \pm 3.86$ and 
$\langle \tau \rangle = 17.14 \pm 0.46$ 
before 1988-11, and 
$\langle T \rangle = 19.71 \pm 4.00$ and 
$\langle \tau \rangle = 17.68 \pm 0.42$ 
after  1988-11. From these results, we observe that the breakpoints are 
associated with a change in the mean value of the trend from these 
time series. Concerning on mean temperature, we observe a change. 
\subsection{Southest Capitals}
\begin{figure}[htbp]
\centering
\includegraphics[scale=.4]{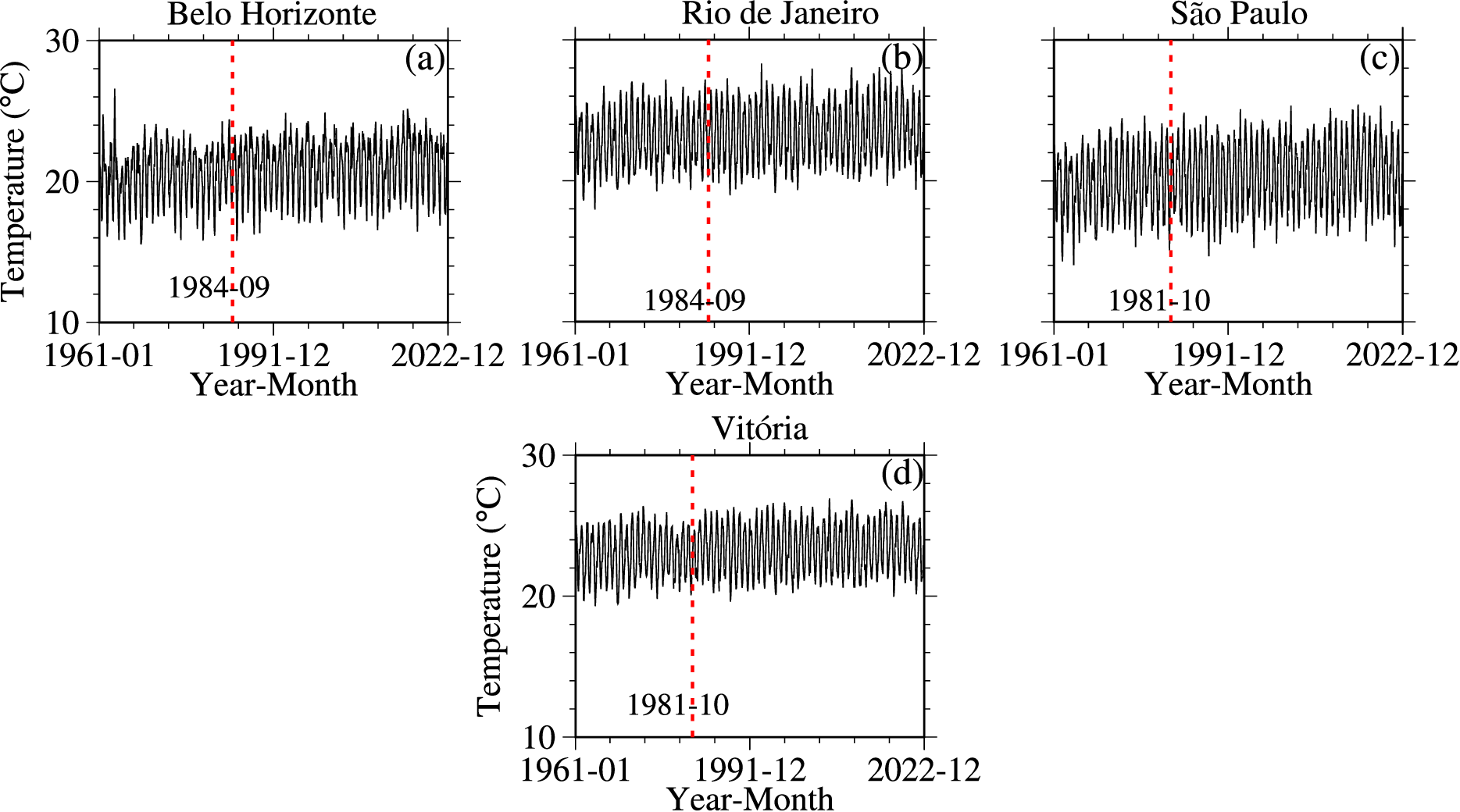}
\caption{Monthly average temperature ($^{\rm o}$C) for the Southeast capitals: 
(a) Belo Horizonte, (b) Rio de Janeiro, (c) S\~ao Paulo, and (d) Vit\'oria. The vertical 
dotted red line marks the breakpoint for each city, with the respective 
mean temperature ($\langle T \rangle$) and trend ($\langle \tau \rangle$) 
before and after the breakpoint.}
\label{fig10}
\end{figure}

Monthly average temperature ($^{\rm o}$C) for the Southeast capitals is 
displayed in Fig. \ref{fig10}. The panels (a), (b), (c), and (d) display 
the results for Belo Horizonte, Rio de Janeiro, S\~ao Paulo, and Vit\'oria, 
respectively. All of the time series are stationary. As observed in the 
results  for the South's capital, the breakpoints occur in 80's 
decade for the Southest Capitals. For Belo Horizonte the breakpoint occurs  
in 1984-09 and before this point the mean values are 
$\langle T \rangle = 20.35 \pm 2.02$ and 
$\langle \tau \rangle = 20.40 \pm 0.50$. After crossing the point the values 
becomes 
$\langle T \rangle = 21.06 \pm 2.05$ and 
$\langle \tau \rangle = 21.06 \pm 0.40$. As observed for the South's capitals, 
a change occurs in the mean temperature. However, due to the high standard 
deviation, we are not able to affirm the change in the temperature. Nonetheless, 
our simulations show a change in the trend. Rio de Janeiro has the same 
breakpoint as Belo Horizonte, i.e., in 1984-09. The average values before 
it are 
$\langle T \rangle = 22.88 \pm 2.03$ and 
$\langle \tau \rangle = 22.88 \pm 0.41$. After the breakpoint are 
$\langle T \rangle = 23.42 \pm 2.15$ and 
$\langle \tau \rangle = 23.42 \pm 0.39$. Rio de Janeiro exhibits temperature 
higher than Belo Horizonte, but the change in the statistical properties 
before and after 1984-09 are in the trend of the time series, which increase. 
S\~ao Paulo also has a similar behaviour, as noted in the results present 
in the panel (c). For example, before 1981-10 the mean values are 
$\langle T \rangle = 19.88 \pm 2.31$ and 
$\langle \tau \rangle = 19.78 \pm 0.46$; and after 1981-10 they change to 
$\langle T \rangle = 20.46 \pm 2.43$ and 
$\langle \tau \rangle = 20.46 \pm 0.40$. The last city in this region is 
Vit\'oria, which has a breakpoint located in 1981-10. Before this point 
we obtain 
$\langle T \rangle = 22.92 \pm 1.77$ and 
$\langle \tau \rangle = 22.92 \pm 0.10$. After the breakpoint we have 
$\langle T \rangle = 23.39 \pm 1.77$ and 
$\langle \tau \rangle = 23.39 \pm 0.10$. The standard deviation 
does not change for $\langle \tau \rangle$, but the $\langle \tau \rangle$ 
value has a considerable change. 
\subsection{Center West}
\begin{figure}[htbp]
\centering
\includegraphics[scale=0.4]{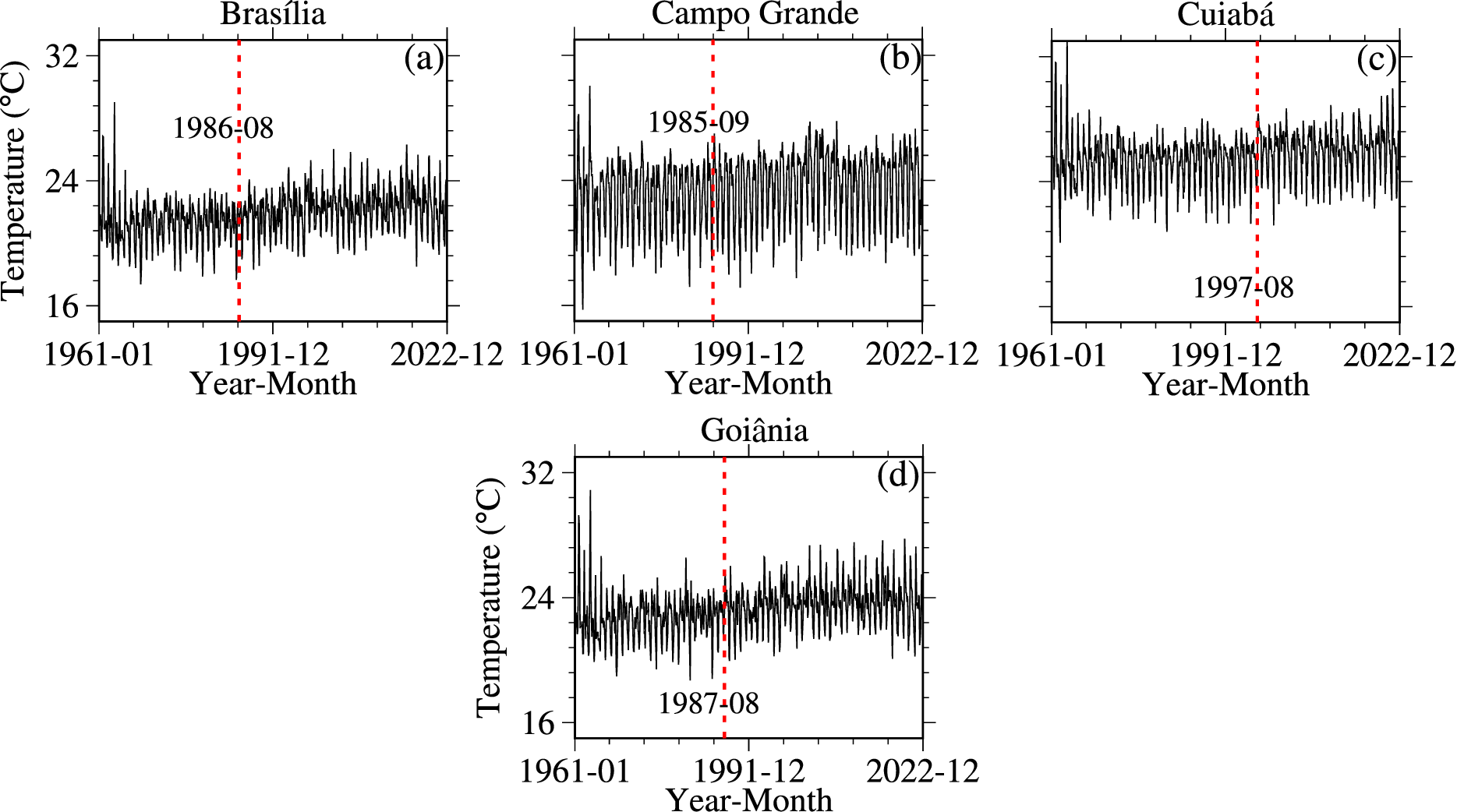}
\caption{Monthly average temperature ($^{\rm o}$C) for the Center West capitals: 
(a) Bras\'ilia, (b) Campo Grande, (c) Cuiab\'a, and (d) Goi\^ania. The vertical 
dotted red line marks the breakpoint for each city, with the respective 
mean temperature ($\langle T \rangle$) and trend ($\langle \tau \rangle$) 
before and after the breakpoint.}
\label{fig11}
\end{figure}

Center West has the capitals Campo Grande, Cuiab\'a, and Goi\^ania as well as  
Federal District with capital named Bras\'ilia. Figure \ref{fig11} 
displays the results for (a) Bras\'ilia, (b) Campo Grande, (c) Cuiab\'a, 
and (d) Goi\^ania. The respective breakpoints occur in 1986-08, 1985-09, 
1997-08, and 1987-08. Considering that the time series are stationary 
for $p$-value $<$ 0.01, all the four cities are stationary. For 
Cuiab\'a, the breakpoint does not occur in 80's decade. Considering 
the results in the panel (a), the values before and after the breakpoint 
are: 
$\langle T \rangle = 21.29 \pm 1.45$ and 
$\langle \tau \rangle = 21.25 \pm 0.33$; and 
$\langle T \rangle = 22.21 \pm 1.38$ and 
$\langle \tau \rangle = 22.21 \pm 0.41$. The standard deviation associated 
with the mean temperature decreases as in relation to the previous results.  
However, it again is not possible a significant change in the temperature. 
On the other hand, we observe an increase in the trend of the time series. 
Before the breakpoint for Campo Grande the statistical values are 
$\langle T \rangle = 22.98 \pm 2.14$ and 
$\langle \tau \rangle = 22.99 \pm 0.27$; after this point they change to 
$\langle T \rangle = 23.75 \pm 2.20$ and 
$\langle \tau \rangle = 23.75 \pm 0.45$. Again the observation is the change 
in the trend of the time series. Cuiab\'a shows a breakpoint in 1997-08. 
Having averages before and after equal to:
$\langle T \rangle = 25.28 \pm 1.59$ and 
$\langle \tau \rangle = 25.27 \pm 0.34$; and 
$\langle T \rangle = 25.95 \pm 1.51$ and 
$\langle \tau \rangle = 25.93 \pm 0.33$. The average temperature does not 
have a significant change on the opposite behaviour of the trend. Finally, 
Goi\^ania has averages equal to 
$\langle T \rangle = 22.70 \pm 1.50$ and 
$\langle \tau \rangle = 22.70 \pm 0.38$ before 1987-08 and 
$\langle T \rangle = 23.63 \pm 1.42$ and 
$\langle \tau \rangle = 23.63 \pm 0.41$ after this point. The values of 
average temperature and trend coincide. The difference is in the standard 
deviation. 
\subsection{North Capitals}
\begin{figure}[htbp]
\centering
\includegraphics[scale=0.4]{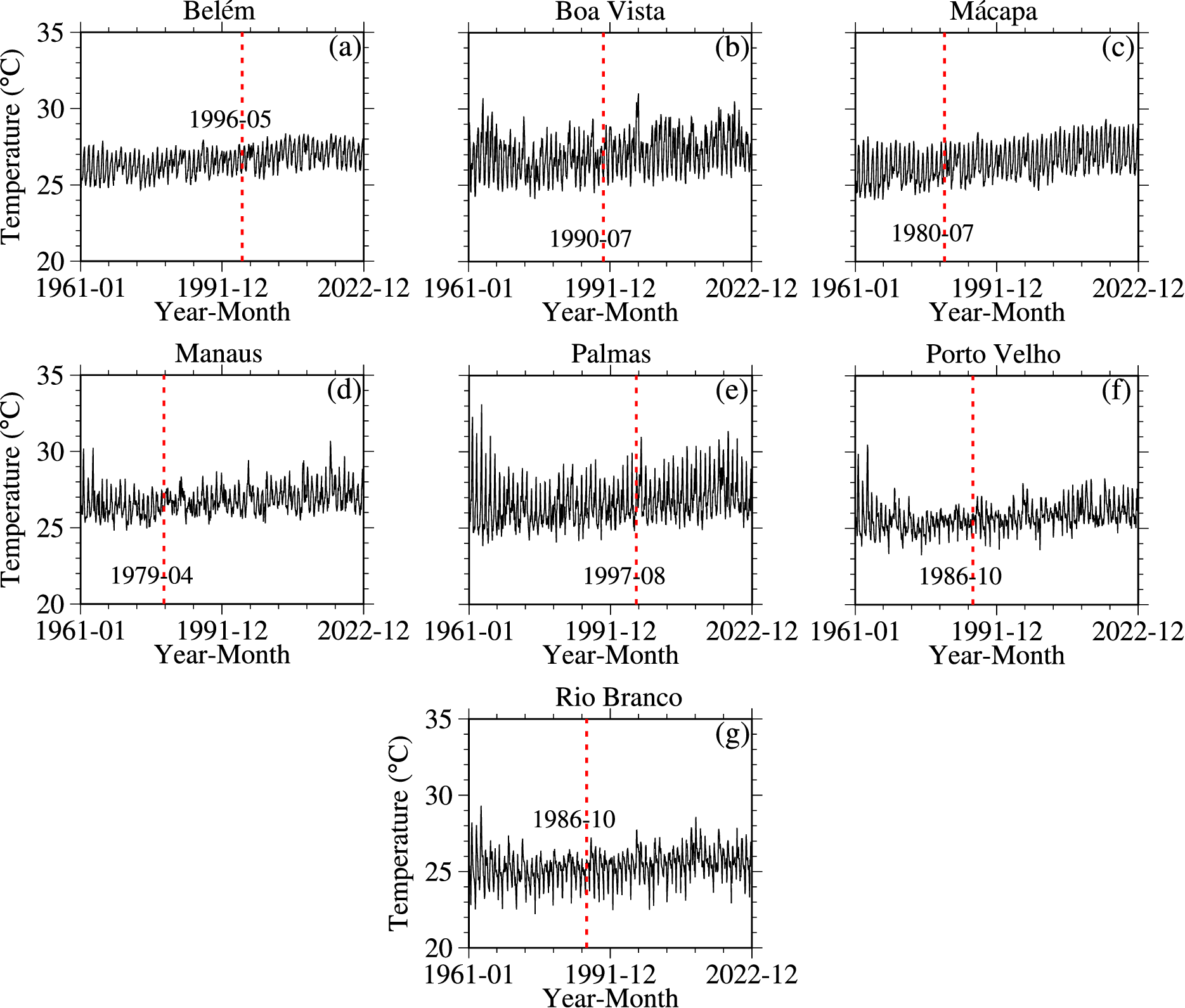}
\caption{Monthly average temperature ($^{\rm o}$C) for the North capitals: 
(a) Bel\'em, (b) Boa Vista, (c) Macap\'a, (d) Manaus, (e) Palmas, (f) Porto Velho, 
and (g) Rio Branco. The vertical 
dotted red line marks the breakpoint for each city, with the respective 
mean temperature ($\langle T \rangle$) and trend ($\langle \tau \rangle$) 
before and after the breakpoint.}
\label{fig12}
\end{figure}

Figure \ref{fig12} shows the time series for (a) Bel\'em, (b) Boa Vista, 
(c) Macap\'a, (d) Manaus, (e) Palmas, (f) Porto Velho, and (g) Rio Branco. 
By the ADF test, our results show that Bel\'em and Macap\'a have non-stationary  
time series. The other cities are stationary. The breakpoint for Bel\'em 
occurs in 1996-05, where the average values before are 
$\langle T \rangle = 26.28 \pm 0.74$ and 
$\langle \tau \rangle = 26.29 \pm 0.28$; and after 
$\langle T \rangle = 27.04 \pm 0.71$ and 
$\langle \tau \rangle = 27.03 \pm 0.26$. Now we observe a significant change 
in the mean value of temperature and trend after the breakpoint. The average 
temperature increases in 0.76 $^{\rm o}$C. Boa Vista has a breakpoint in 
1990-07 with averages before it equal to 
$\langle T \rangle = 26.64 \pm 1.25$ and 
$\langle \tau \rangle = 26.64 \pm 0.42$; and after 
$\langle T \rangle = 27.38 \pm 1.23$ and 
$\langle \tau \rangle = 27.39 \pm 0.44$. Different from Bel\'em, the change 
after 1990-07 occurs only in trend. Macap\'a also presents a significant 
change only in the trend after its breakpoint  in 1980-07. The 
average values before and after this data are, respectively: 
$\langle T \rangle = 25.90 \pm 1.05$ and 
$\langle \tau \rangle = 25.93 \pm 0.26$;  
$\langle T \rangle = 26.81 \pm 1.04$ and 
$\langle \tau \rangle = 26.81 \pm 0.42$. 
Manaus has the breakpoint in 1979-04, with statistical values equal to 
$\langle T \rangle = 26.37 \pm 0.86$ and 
$\langle \tau \rangle = 26.36 \pm 0.28$ before; and 
$\langle T \rangle = 26.95 \pm 0.82$ and 
$\langle \tau \rangle = 26.94 \pm 0.38$ after. For Manaus, a relative small 
change occurs in the average trend. The panel (e) exhibits the results for 
Palmas, where the breakpoint occurs in 1997-08. For Palmas, the averages 
before the breakpoint are 
$\langle T \rangle = 26.26 \pm 1.34$ and 
$\langle \tau \rangle = 26.36 \pm 0.35$; and after: 
$\langle T \rangle = 27.12 \pm 1.33$ and 
$\langle \tau \rangle = 27.11 \pm 0.43$. The significance change is 
 in trend. Porto Velho shows a standard deviation in average temperature 
less than one. However, it is not possible  a significant change 
in this value after the breakpoint, occuring in 1986-10. The value 
in mean temperature goes from $\langle T \rangle = 25.41 \pm 0.92$ 
to $\langle T \rangle = 25.96 \pm 0.79$. However, in the mean trend 
the change is significant. Before 1986-10, our simulations suggest 
$\langle \tau \rangle = 25.40 \pm 0.36$ and 
$\langle \tau \rangle = 25.92 \pm 0.35$. The last analysed city in the 
North region is Rio Branco. The breakpoint in Rio Branco occurs in 1986-10 
and the average temperature does not change at all. Before, this point is 
$\langle T \rangle = 25.08 \pm 1.00$ and after is $\langle T \rangle = 25.57 \pm 0.95$. 
The significant change happens in the average trend, that goes from 
$\langle \tau \rangle = 25.08 \pm 0.33$ to $\langle \tau \rangle = 25.57 \pm 0.37$. 
\subsection{Northeast Capitals}
\begin{figure}[htbp]
\centering
\includegraphics[scale=0.4]{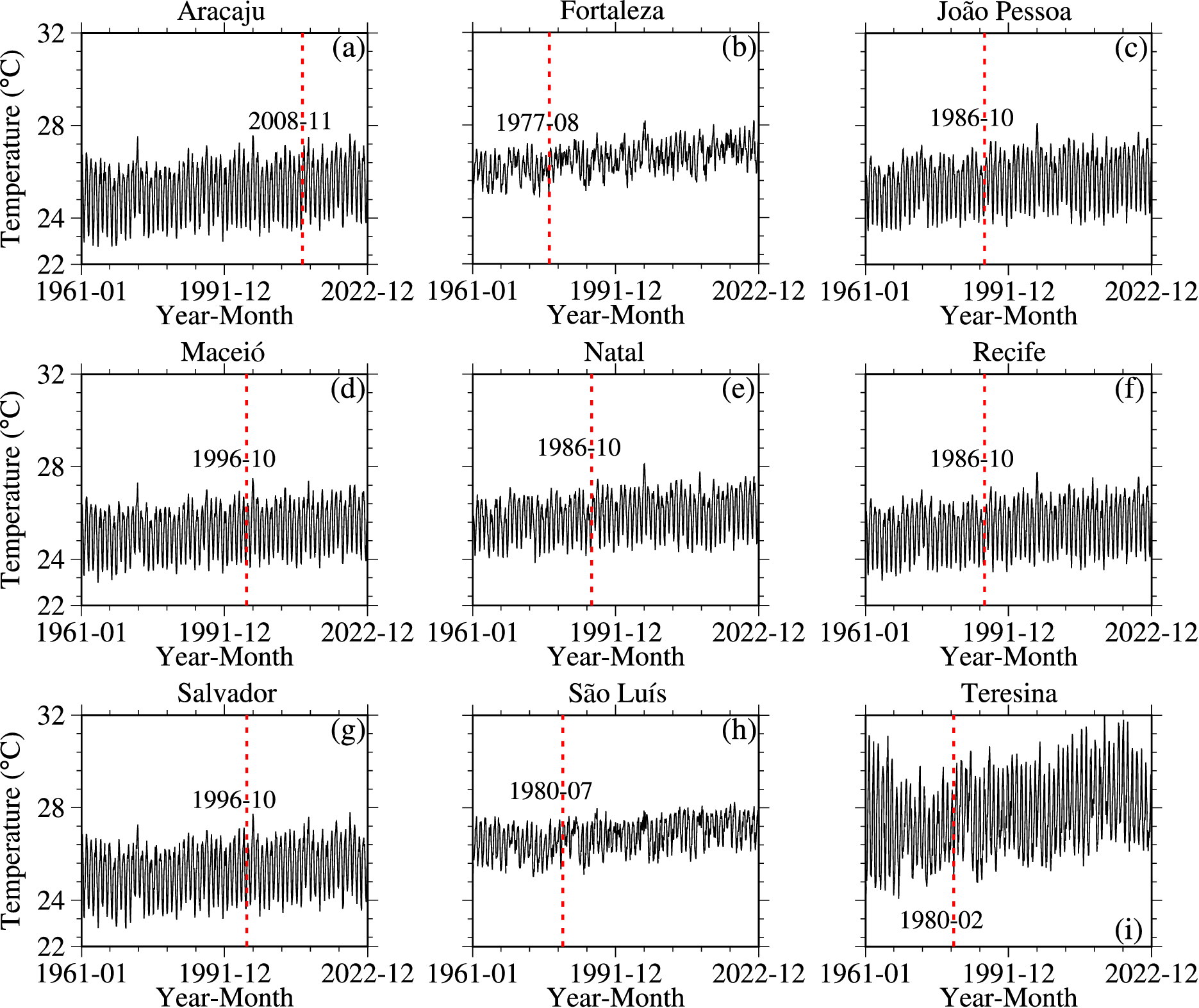}
\caption{Monthly average temperature ($^{\rm o}$C) for the Northeast capitals: 
(a) Aracaju, 
(b) Fortaleza, 
(c) Jo\~ao Pessoa, 
(d) Macei\'o, 
(e) Natal, 
(f) Recife, 
(g) Salvador, 
(h) S\~ao Lu\'is, and 
(i) Teresina. The vertical 
dotted red line marks the breakpoint for each city, with the respective 
mean temperature ($\langle T \rangle$) and trend ($\langle \tau \rangle$) 
before and after the breakpoint.}
\label{fig13}
\end{figure}

The last considered Brazilian region is Northeast. Figure \ref{fig13} 
exhibits the results for the North Capitals: 
(a) Aracaju, 
(b) Fortaleza, 
(c) Jo\~ao Pessoa, 
(d) Macei\'o, 
(e) Natal, 
(f) Recife, 
(g) Salvador, 
(h) S\~ao Lu\'is, and 
(i) Teresina. These 9 time series are stationary. Our simulations show  
that Aracaju has a breakpoint in 2008-11. Before this point the averages 
are  
$\langle T \rangle = 25.19 \pm 1.10$ and 
$\langle \tau \rangle = 25.19 \pm 0.26$;  
$\langle T \rangle = 25.64 \pm 1.07$ and after are  
$\langle \tau \rangle = 25.65 \pm 0.21$. It occurs an increase in the 
average temperature that is not significant due to the high standard deviation. 
On the other hand, the change in the trend it is noticeable. Fortaleza 
exhibits a signification change in the average temperature when compared 
before and after the breakpoint, which occurs in 1977-08. The average 
temperature goes from $\langle T \rangle = 26.07 \pm 0.55$ to 
$\langle T \rangle = 26.70 \pm 0.56$. Another change that occurs is in 
the trend, that goes from $\langle \tau \rangle = 26.09 \pm 0.22$ to 
$\langle \tau \rangle = 26.70 \pm 0.31$. Jo\~ao Pessoa shows a breakpoint 
in 1986-10. The average temperature does not change significantly, e.g., 
before its value is $\langle T \rangle = 25.41 \pm 0.96$ and after is 
$\langle T \rangle = 25.85 \pm 1.04$. A small change occurs in the average 
trend value, that goes from $\langle \tau \rangle = 25.42 \pm 0.23$ 
to $\langle \tau \rangle = 25.85 \pm 0.22$. Macei\'o has its breakpoint 
in 1996-10. The average values before are 
$\langle T \rangle = 25.19 \pm 0.96$ and 
$\langle \tau \rangle = 25.20 \pm 0.21$;  
$\langle T \rangle = 25.57 \pm 0.97$ and after are  
$\langle \tau \rangle = 25.56 \pm 0.20$. Natal also has a similar behaviour. 
The difference is that the breakpoint for Natal occurs in 1986-10 and the 
associated averages before and after are 
$\langle T \rangle = 25.70 \pm 0.79$ and 
$\langle \tau \rangle = 25.71 \pm 0.21$;  
$\langle T \rangle = 26.11 \pm 0.88$ and   
$\langle \tau \rangle = 26.11 \pm 0.22$, respectively. Our analyses show  
that Recife has a breakpoint in the same point that Natal. The values 
are 
$\langle T \rangle = 25.19 \pm 0.96$ and 
$\langle \tau \rangle = 25.20 \pm 0.20$ before 1986-10;  
$\langle T \rangle = 25.60 \pm 1.01$ and   
$\langle \tau \rangle = 26.60 \pm 0.21$ after 1986-10. Salvador has the 
breakpoint located in 1996-10 which split the averages from 
$\langle T \rangle = 25.16 \pm 1.07$ and 
$\langle \tau \rangle = 25.16 \pm 0.24$; to   
$\langle T \rangle = 25.59 \pm 1.05$ and   
$\langle \tau \rangle = 25.59 \pm 0.24$. There is no a significant change 
in the temperature, the change is in the trend. S\~ao Lu\'is has a 
similar behaviour as Salvador. The difference is that the breakpoint 
is located in 1980-07 and the averages values are 
$\langle T \rangle = 26.41 \pm 0.65$ and 
$\langle \tau \rangle = 26.43 \pm 0.20$ before and then change to 
$\langle T \rangle = 26.96 \pm 0.64$ and 
$\langle \tau \rangle = 26.96 \pm 0.30$ after 1980-07, showing a change 
only in the trend. The last city is Teresina. Teresina shows a huge 
variations in the amplitude when compared with the previous Northeast's cities. 
In addition, the breakpoint for Teresina occurs in 1980-02 separating 
the averages values from 
$\langle T \rangle = 27.03 \pm 1.65$ and 
$\langle \tau \rangle = 27.06 \pm 0.41$  to 
$\langle T \rangle = 27.98 \pm 1.57$ and 
$\langle \tau \rangle = 27.98 \pm 0.56$ showing a considerable change only 
in the trend.     
\newpage
\section{Appendix B}\label{apendiceb}
To evaluate the performance of the forecasting, we calculate the absolute error 
by
\begin{equation}
\Delta {\rm E} = |T_k - \widehat{T}_k|,
\label{eq_erro}
\end{equation}
where $T_k$ is the real data and $\widehat{T}_k$ 
is the predicted one. We also  compute 
the root mean square error (RMSE), by 
\begin{equation}
{\rm RMSE} = \sqrt{\frac{\sum_{k=1}^{N} (T_k - \widehat{T}_k)^2}{N}}, \label{rmse}
\end{equation}
the correlation coefficient:  
\begin{equation}
r = \frac{\sum_{k=1}^{N} (T_k - \langle{T}\rangle) (\widehat{T}_k - \langle{\widehat{T}\rangle)}}{\sqrt{\sum_{k=j}^{N} (T_k - \langle{T}\rangle)^2 \sum_{k=j}^N (\widehat{T_k} - \langle{\widehat{T}\rangle)^2}}},
\label{r_eq}
\end{equation}
{
the mean absolute error (MAE) \cite{Willmott2023}:
\begin{equation}
{\rm MAE} = \frac{1}{N} \sum_{k=1}^N |T_k - \widehat{T}_k|, \label{mae}
\end{equation}
the normalized root mean squared error (NRMSE) \cite{Jadon2024}:
\begin{equation}
{\rm NRMSE} = \frac{1}{\rho}\sqrt{\frac{1}{N} \sum_{k=1}^{N} (T_k - \widehat{T}_k)^2}, \label{nrmse}
\end{equation}
where $\rho$ is the normalization constant, which can be 
$\rho = \langle{T}\rangle$ (average of temperature), $\rho=T_{\rm max} - T_{\rm min}$, or the standard 
deviation of $T$. We use the first normalization for normal data and the 
second for anomalous one, once the average for anomalous temperature is small 
value and can be negative in certain cases.

Additionaly, we calculate the refined Willmott's index \cite{Willmott2012}, defined by 
\begin{equation}
d_r =
\left\{
\begin{array}{ll}
1 - \frac{ \sum_{k=1}^{N} |\widehat{T}_k - T_k| }{ c \sum_{k=1}^{N} |T_k - \langle{T}\rangle| }, & \text{if} \quad \sum_{k=1}^{N} |\widehat{T}_k - T_k| \leq c \sum_{k=1}^{N} |T_k - \langle{T}\rangle| \\[12pt]
\\
\frac{c \sum_{k=1}^{N} |T_k - \langle{T}\rangle|}{ \sum_{k=1}^{N} |\widehat{T}_k - T_k| } - 1, & \text{if} \quad \sum_{k=1}^{N} |\widehat{T}_k - T_k| > c \sum_{k=1}^{N} |T_k - \langle{T}\rangle|, \label{willmott}
\end{array}
\right.
\end{equation}
where $c=2$ \cite{Willmott2012}. This index measures the 
magnitudes of difference between the predicted ($\widehat{T}_k$) and 
observed ($T_k$) values. For instance, if $d_r = 0.5$, the sum of the error is 
one half of the sum of the perfect model ($\widehat{T}_k =T_k$) deviation.}

The notation $\langle{T}\rangle$ and $\langle{\widehat{T}}\rangle$ means the averages 
of real and predicted temperatures, respectively, in the considered time window, 
defined from $j=1,2,3,...,N-1$ until $N$. 
\end{document}